\documentclass[11pt]{article}

\usepackage{amsfonts,nicefrac,bbm}
\usepackage{amsmath}
\usepackage{amssymb}
\usepackage[round]{natbib}
\usepackage[T1]{fontenc}
\usepackage{graphicx}
\usepackage{booktabs}
\usepackage{geometry, tabularx}
\usepackage{booktabs}
\usepackage{float}
\usepackage{multirow}
\usepackage{caption}
\usepackage{subfig}
\usepackage{amsmath} 
\usepackage{makecell}
\usepackage{extarrows}
\usepackage{enumitem}

\providecommand{\U}[1]{\protect\rule{.1in}{.1in}}

\newtheorem{theorem}{Theorem}[section]

\newtheorem{corollary}[theorem]{Corollary}

\newtheorem{definition}[theorem]{Definition}
\newtheorem{example}[theorem]{Example}

\newtheorem{lemma}[theorem]{Lemma}
\newtheorem{assumption}[theorem]{Assumption}

\newtheorem{problem}[theorem]{Problem}
\newtheorem{proposition}[theorem]{Proposition}
\newtheorem{remark}[theorem]{Remark}

\newcommand{\p}{\mathbb{P}}

\newcommand{\cIr}{\mathcal{IR}}

\newcommand{\cIc}{\mathcal{IC}}
\newcommand{\E}{\mathbb{E}}
\newcommand{\bbR}{\mathbb{R}}

\renewcommand{\tilde}{\widetilde}

\DeclareMathOperator*{\argmax}{\arg\max}

\newenvironment{proof}[1][Proof]{\noindent\textbf{#1.} }{\ \rule{0.5em}{0.5em}}
\newenvironment{proof*}{\par\noindent}{\hfill$\square$\par}

\numberwithin{equation}{section}
\allowdisplaybreaks
\newcommand{\One}{\mathbbm{1}}

\usepackage[table,xcdraw]{xcolor}
\usepackage{xr-hyper}
\definecolor{darkblue}{rgb}{0.1,0.1,0.9}
\definecolor{darkred}{rgb}{0.8,0.1,0.1}
\usepackage[colorlinks=true, allcolors=darkblue, urlcolor=darkred]{hyperref}

\allowdisplaybreaks

\begin{document}
\title{Optimal Insurance in a Monopoly:\vspace{0.15cm}\\Dual Utilities with Hidden Risk Attitudes}

\author{Mario Ghossoub\,\thanks{Department of Statistics and Actuarial Science, University of Waterloo (email: \href{mailto:mario.ghossoub@uwaterloo.ca}{mario.ghossoub@uwaterloo.ca}).\vspace{0.1cm}}
\and  
Bin Li\,\thanks{Department of Statistics and Actuarial Science, University of Waterloo (email: \href{mailto:bin.li@uwaterloo.ca}{bin.li@uwaterloo.ca}).\vspace{0.1cm}}
\and 
Benxuan Shi\,\thanks{Department of Statistics and Actuarial Science, University of Waterloo (email: \href{mailto:benxuan.shi1@uwaterloo.ca}{benxuan.shi1@uwaterloo.ca}).}}


\maketitle
\vspace{-.25in}

\begin{abstract}
We consider a monopoly insurance market with a risk-neutral profit-maximizing insurer and a consumer with Yaari Dual Utility preferences that distort the given continuous loss distribution. The insurer observes the loss distribution but not the risk attitude of the consumer, proxied by a distortion function drawn from a continuum of types. We characterize the profit-maximizing, incentive-compatible, and individually rational menus of insurance contracts, show that equilibria are separating, and provide key properties thereof. Notably, insurance coverage and premia are monotone in the level of risk aversion; the most risk-averse consumer receives full insurance (\textit{efficiency at the top}); the monopoly absorbs all surplus from the least-risk averse consumer; and consumers with a higher level of risk aversion induce a higher expected profit for the insurer. Under certain regularity conditions, equilibrium contracts can be characterized in terms of the marginal loss retention per type of consumer, and they consist of menus of layered deductible contracts, where each such layered structure is determined by the risk type of the consumer. In addition, we examine the effect of a fixed insurance provision cost on equilibria. We show that if the fixed cost is prohibitively high, then there will be no \textit{ex ante} gains from trade. However, when trade occurs, separating equilibrium contracts always outperform pooling equilibrium contracts, and they are identical to those obtained in the absence of fixed costs, with the exception that only part of the menu is excluded. The excluded contracts are those designed for consumers with relatively lower risk aversion, who are less valuable to the insurer. Finally, we characterize incentive-efficient menus of contracts in the context of an arbitrary type space.
\end{abstract}



\vskip12pt
\noindent\textbf{Acknowledgments:}
Mario Ghossoub acknowledges financial support from the Natural Sciences and Engineering Research Council of Canada (NSERC Grant No.\ 2024-03744). Bin Li acknowledges financial support from (NSERC Grant No.\ 2020-04338). Benxuan Shi acknowledges financial support from the Society of Actuaries through the Hickman Scholars Program.

\newpage
\section{Introduction}\vspace{-0.25cm}

The theoretical foundations of the economics of risk and insurance were laid out by \cite{Borch1962a}, \cite{Arrow1963b, Arrow1971}, and \cite{Wilson1968}, upon the framework provided by the general equilibrium theory under uncertainty of \cite{allais53b}, \cite{Arrow1953}, and \cite{debreu59}. This allowed an approach to risk-sharing markets based on the view that risks can be traded through markets for state-contingent goods, assumed to be perfectly competitive and complete, and hence to relate price-mediated risk-exchange markets to welfare economics. However, a major lacuna in this edifice, initially remarked upon by \cite{allais53b} and \cite{Arrow1963b} themselves, is the fact that these risk-sharing markets are both incomplete and imperfect in practice. Major sources of incompleteness and imperfection in these markets are the uninsurability of some sources of risk, as \cite{Arrow1963b} noted, and the asymmetry in information, as \cite{allais53b} recounted. The latter stems either from moral hazard (\textit{ex ante} or \textit{ex post}), or from adverse selection. Other sources of market imperfection are also outlined in \cite{Arrow1963b}.

Insurers have long been concerned with the effect of imperfect information on contractual agreements, premium calculations, and profitability. Yet, despite the warnings of \cite{allais53b} and  \cite{Arrow1963b}, the initial theoretical strides in the study of insurance markets conveniently ignored information problems, until the seminal contributions of \cite{RothschildandStiglitz1976} and \cite{stiglitz1977} on the problem of adverse selection; \cite{Pauly74}, \cite{Marshall76}, and \cite{Shavell79} on the problem of \textit{ex ante} moral hazard; and \cite{SpenceZeckhauser71} and \cite{Townsend79}
 on the problem of \textit{ex post} moral hazard.

Our focus in this paper is on the literature on hidden types and adverse selection in insurance markets. The first systematic study of these issues was carried out by \cite{RothschildandStiglitz1976} in the context of competitive insurance markets with risk-neutral insurers. The authors highlighted some key differences between competitive markets with perfect information and competitive markets with adverse selection. In the former case, equilibria are reached and they induce full insurance coverage at own actuarial odds for each consumer, with zero profit for insurers. Moreover, individuals strictly prefer insurance to no insurance, thereby leading to some consumer surplus in the market not being extracted by insurers. In the case of adverse selection in competitive markets (that is, if consumers are of either high-risk or low-risk type, differ only in their risk distribution but not in their attitude toward risk, and their risk types are unobservable by the insurers), \cite{RothschildandStiglitz1976} showed that  the market does not necessarily reach a (Cournot-Nash) equilibrium; but if it does, then this equilibrium must be a separating equilibrium, whereby each risk type receives a different type of coverage, and each equilibrium contract induces zero profit to insurers. Specifically, the separating equilibrium provides full insurance at a high price to the high-risk types and only partial insurance at a lower price to the low-risk types. However, both types strictly prefer some form of insurance to no insurance at all. Hence, the consumer surplus is not all absorbed by insurers. Nonetheless, in this case, the high-risk types cause a negative externality to the low-risk types, in that equilibrium contracts induce a welfare loss to the low-risk types compared to the perfect information case: equilibrium contracts yield the same expected utility to the high-risk types as in the perfect information case, but they lead to a lower expected utility for the low-risk types. Hence, these separating equilibria may not be Pareto optimal. These findings were extended by \cite{YoungandBrowne2000} to the case of the Dual Utility (DU) model of \cite{yaari}, in which the attitude to risk is captured by probability weighting rather than marginal utility of wealth.  

The case of a risk-neutral monopoly with no transaction costs was first examined by \cite{stiglitz1977}, where two types of consumers differ with regard to their loss distribution, but not with respect to their attitude toward risk, that is, their marginal utility of wealth. With perfect information, equilibria consist of full-insurance contracts at premia that make the consumers indifferent between purchasing insurance and foregoing coverage. That is, all the consumer surplus involved in the reduction of risk is extracted by the monopoly, as is classically understood. When risk types are hidden information to the insurer, \cite{stiglitz1977} highlighted some important differences between competitive and monopoly equilibria. In particular, there are always \textit{ex ante} gains from risk sharing, monopoly equilibria are separating, and the high-risk consumers always purchase full insurance. However, the low-risk consumers may prefer not to purchase any coverage. When they do, they receive only partial coverage but no consumer surplus.

Another key contribution of \cite{stiglitz1977} was to show that, while in competitive insurance markets only differences in risk distribution affect equilibria and the nature of separating contracts, in a monopoly both risk and risk attitude have an impact on equilibrium contracts. In particular, if the market were competitive, then risk-neutral competitive insurers offer full insurance, regardless of the consumers' levels of risk aversion. In a monopoly, however, this information can be valuable. Indeed, \cite{stiglitz1977} showed that if consumers are of two types, namely highly risk averse and mildly risk averse, and if the monopolist knew the consumers' level of risk aversion, then equilibria entail full insurance to each at terms that make each type of consumer indifferent between purchasing insurance and not doing so. However, if the consumer's level of risk aversion is hidden from the insurer, then equilibrium contracts can have different forms, and they can be separating or pooling equilibria.

The insight gained from the pioneering work of  \cite{stiglitz1977} spawned a large literature on monopoly in insurance markets with hidden types, be it hidden risk distributions or hidden risk attitudes. Notably, \cite{chade2012} extend the model of \cite{stiglitz1977} to account for a continuum of consumer types. They show that all of the results of \cite{stiglitz1977} still hold, and in particular that, in equilibrium, the insurer expects a strictly positive profit, the highest risk type receives full insurance, coverage is monotone in types, all other types receives partial insurance, and the lowest type is indifferent between insurance and no insurance. Recently, \cite{gershkov2023optimal} proposed an extension of the model of \cite{stiglitz1977} in a different direction. They consider a monopoly market with a risk-neutral insurer and a continuum of consumer types, each of which is subject to a loss with a continuous distribution. The loss distribution is hidden from the insurer, and the consumers' riskiness is ordered by first-order stochastic dominance of the loss distributions. The consumers are assumed to behave according to the DU model of \cite{yaari}, with a common distortion function representing risk attitude and known by the insurer. They show that under a regularity condition, the optimal coverage is a layered deductible indemnity schedule. Moreover, under some technical conditions, they provide sufficient conditions for the optimal menus to consist either of linear deductible contracts or of upper limit contracts.
Another recent extension of the classical model is due to \cite{CHADE2020105085}, who propose to study the effect of insurance provision costs (namely costs that are in addition to the indemnification costs) on the predictions of the classical model of \cite{stiglitz1977} regarding the properties of equilibria in monopoly markets with hidden risk distributions: there are always \textit{ex ante} gains from risk sharing; monopoly equilibria are separating; the high-risk consumers always purchase full insurance; the low-risk consumers may prefer not to purchase any coverage;  when they do purchase coverage, the low-risk consumers receive only partial coverage but no consumer surplus. \cite{CHADE2020105085} show that all of these classical properties might fail to hold in the presence of insurance provision costs. Namely, they show that two widely popular types of provision costs can lead to no trade, and hence market failure. When there is trade, they show how insurance costs can lead to pooling equilibria, as well as a breakdown in the predictions of the classical model.

Motivated by the observation of \cite{stiglitz1977} that in a monopoly both risk and risk attitude have an impact on equilibrium contracts, \cite{LANDSBERGER1994392} consider a monopoly market with a risk-neutral insurer and two types of EU consumers, differing only in their level of risk aversion, not in their risk distribution. The level of risk aversion is hidden from the insurer, while the loss distribution is known. They show that, under a limited liability constraint on the individuals' risk exposure, separating equilibria are such that the more risk averse types receive the certainty equivalent of their loss random variable, and the less risk averse types receive two-valued partial insurance. Moreover, the more risk averse types do not receive any consumer surplus: they are indifferent between insurance and no insurance. In a similar vein, \cite{Boonen_Zhang_2021} consider a monopoly market with a risk-neutral insurer and two types of consumers, whose preferences are assumed to be represented by distortion risk measures. These are similar mathematical objects to the Dual Utilities of \cite{yaari}. The consumers are subject to a loss with an arbitrary continuous distribution, which is known by the insurer. However, the risk preferences of the consumers, proxied by their distortion functions, is hidden from the insurer. Specifically, the authors assume that the consumers are ranked by their willingness to pay to eliminate risk exposure. Equivalently, in the model of \cite{yaari}, this amounts to a ranking based on pessimism, or weak risk aversion (see \cite{GhossoubHe2021}). The authors examine both separating and pooling equilibria. They find that in a separating equilibrium, the more pessimistic type receives full insurance and can capture some consumer surplus, whereas the less pessimistic type is offered only partial insurance and is indifferent between purchasing and not purchasing insurance coverage. In a pooling equilibrium, the monopoly receives a lower net profit compared to a separating equilibrium. Moreover, in a pooling equilibrium, the less pessimistic type receives no consumer surplus and is therefore indifferent between insurance and no insurance. 

In this paper, we examine the case of a monopoly insurance market with hidden risk attitudes. Specifically, the market is populated by a risk-neutral profit-maximizing insurer and a consumer with DU preferences, as in \cite{yaari}. The consumer is facing an insurable loss with a continuous distribution function, which is known by the insurer. However, the insurer cannot observe the risk attitude of the weakly risk-averse consumer, which is proxied by a distortion function that is drawn from a distribution over a continuum of types. We characterize the optimal (profit-maximizing) incentive-compatible and individually rational menus of insurance contracts. We find that, under certain regularity conditions, an optimal such menu can be characterized in terms of the marginal loss retention per type of consumer, and it consists of a layered deductible contract, where each such layered structure is determined by the risk type of the consumer. Such layered indemnity structures are widely observed in practice. In addition, we show that a separating equilibrium is always more profitable to the insurer than a pooling equilibrium, and it induces a strictly positive expected profit to the insurer. This extends the key findings of \cite{Boonen_Zhang_2021} to a setting of continuum consumer types, and it echoes similar results in the case of common risk attitudes but unobservable risk distributions.

The key properties of our separating equilibria are the following. First, insurance coverage and premia are monotone in the level of risk aversion. Second, the most risk-averse consumer receives full insurance (\textit{efficiency at the top}). Third, the monopoly will absorb all surplus from the least-risk averse consumer, and the contract offered to this consumer leaves them indifferent between participation and non-participation. Fourth, we elicit an interesting monotonicity property of the consumer's utility of wealth at an optimum, as a function of their level of risk aversion. Fifth, consumers with a higher level of risk aversion, who are willing to pay more for a product with a higher coverage, are generally more valuable to the insurer, in that they induce a higher expected profit for the insurer. This contradicts the findings in scenarios with a hidden loss distribution, such as in \cite{chade2012} and \cite{gershkov2023optimal}, where the most profitable consumers are always of intermediate, rather than extreme, types. Finally, we find that with perfect information, that is, when the risk attitude of the consumer is observed by the monopoly, equilibria entail full-insurance at premia that make the consumer indifferent between purchasing insurance and foregoing coverage, thereby extending one of the classical findings of \cite{stiglitz1977}.

In addition, and following \cite{CHADE2020105085}, we examine the effect of insurance provision costs on equilibria in our monopoly market with DU consumer, when risk attitudes are unobservable but risk distributions are. Specifically, we assume that in addition to the expected insurance indemnification payment, the insurer incurs ``friction costs'', such as claims processing costs, administrative expenses, actuarial loading, etc. We consider the case of a fixed cost that occurs only when insurance is provided. We show that the optimal menu is the same as in the absence of such  fixed costs, with the exception that only part of the menu is excluded. The exclusion occurs from below, with the excluded contracts being those typically designed for consumers with relatively lower risk aversion, who are less valuable to the insurer. This effect is also discussed in \citet[Theorem 3 and Corollary 1]{CHADE2020105085}, which shows that when a fixed cost is included, only pooling contracts and no-coverage contracts, both with a positive measure, remain. Moreover, we show that if the fixed cost is prohibitively high, exceeding the profit that the insurer could have obtained in a market with perfect information and no friction cost, and in which the consumer is of the most risk-averse type, then the insurer anticipates losses even on the most risk-averse consumers (and hence on all other types). In that case, there will be no gains from trade, and the insurer is not willing to offer any contract, thereby echoing a similar result of \cite{CHADE2020105085}. However, when trade occurs, separating equilibrium contracts always outperform pooling equilibrium contracts, contrary to \cite{CHADE2020105085}.

Finally, we characterize incentive-efficient menus of contracts in the context of an arbitrary type space. We show that individually rational and incentive compatible contracts that are Pareto optimal can be achieved by maximizing a social welfare function that accounts for hidden types. While it is difficult to solve such a problem in the general case, we are able to provide a crisp characterization of solutions under a few assumptions.

The rest of this paper is structured as follows. Section \ref{SecSetting} presents our market model and formulates the problem. Section \ref{SecIRIC} characterizes the set of individually rational and incentive compatible menus of contracts. Section \ref{Profit-O-menu} characterizes optimal menus and separating equilibria, and examines properties thereof. In Section \ref{SecCost} we study the effects of friction costs on the structure of equilibria, and in Section \ref{SecPO} we examine Pareto-optimal menus of contracts. Finally, Section \ref{SecConc} concludes. All proofs are relegated to the \hyperlink{LinkToAppendix}{Appendix}.

\section{The Insurance Market}\vspace{-0.25cm}
\label{SecSetting}

\subsection{Setup}\vspace{-0.25cm}
We consider a one-period monopoly insurance market in which an consumer is subject to an insurable loss and seeks insurance coverage from a monopolistic insurer, in exchange for a premium payment. We assume that the loss random variable $L$ is an element of $B(\Sigma)$, the space of bounded and $\Sigma$-measurable real-valued functions on a given probability space $(S, \Sigma, \p)$, with range $[0,\overline{L}]$, for some $\overline{L} < +\infty$, and cumulative distribution function $H(l) := \mathbb{P}\left(L \leq l\right)$.

The monopolistic insurer is assumed to be a risk-neutral Expected-Utility maximizer, who aims to maximize profit. The consumer behaves according to the Dual Utility (DU) framework of \cite{yaari}, in which risk aversion is entirely captured by the consumer's probability weighting function, also referred to as a distortion function. 

\begin{definition} 
A function $g:[0,1]\rightarrow[0,1]$ is called a distortion function if it satisfies the following conditions: 
\begin{enumerate}
\item $g(0)=0$ and $g(1)=1$;
\item $g$ is increasing on $[0,1]$.
\end{enumerate}
\end{definition}

Let $X$ be a bounded random variable. The dual utility of $X$ is  the Choquet integral of $X$ with respect to the distorted probability measure $g \circ \mathbb{P}$, given by
\begin{equation}
\label{yaari}
\begin{split}
DU(X) = \int \, X \, \mathrm{d} g\circ \mathbb{P} 
&:= \int_0^{+\infty}g(1-\mathbb{P}(X\leq x))\,\mathrm{d}x+\int_{-\infty}^0
\left[g(1-\mathbb{P}(X\leq x))-1\right]\mathrm{d}x\\
&= - \int_{\mathbb{R}}\,x \, \mathrm{d} g\left(1-F_{X}(x)\right)\\
\end{split}
\end{equation}

\noindent By basic properties of the Choquet integral (e.g., \cite{Denneberg94}), $DU$ is monotone, positively homogeneous, and translation-invariant. In particular, 
$$DU(X+c) = DU(X) + c,$$
for any constant $c \in \mathbb{R}$.

The consumer can purchase insurance coverage $I(L)$ against $L$, satisfying $0 \leq I(L) \leq L$, from the insurer at a premium $p$. The consumer's resulting end-of-period wealth is given by
$$-p - L + I(L) = -p - R(L),$$
where $R(L) = L - I(L) \geq 0$ is the part of the loss that is not covered by the insurer, and hence retained by the consumer. We refer to $R(L)$ as the retained loss. The resulting utility of the consumer is then given by
\begin{align*}
DU(-p - R(L))
&= -p + DU(-R(L))\\
&= -p + \int_{-\infty}^0
\left[g(1-\mathbb{P}(-R(L)\leq x))-1\right]\mathrm{d}x\nonumber\\
&=-p + \int_{-\infty}^0
\left[g(\mathbb{P}(R(L)\leq -x))-1\right]\mathrm{d}x\\
&=-p  -\int_0^{+\infty}
\left[1-g(\mathbb{P}(R(L)\leq l))\right]\mathrm{d}l.\\
&=-p  -\int_0^{\overline{L}}
\left[1-g(\mathbb{P}(R(L)\leq l))\right]\mathrm{d}l.
\end{align*}

\noindent We use the convention that $U(L,0)$ denotes the utility of wealth in the absence of insurance. That is,
$$U(L,0) = DU(-L) = -\int_0^{{\overline{L}}}
\left[1-g(H(l))\right]\mathrm{d}l.$$

By a classical result of \cite{quiggin93} and \cite{ChateauneufCohen1994}, the functional $DU$ in \eqref{yaari} is weakly risk averse if and only if $g(x) \leq x$, for all $x\in [0,1]$. See \cite{GhossoubHe2021} for more about risk aversion in Rank-Dependent Utility Theory and Dual Utility Theory. We make this assumption all throughout.

\begin{assumption}[Risk Aversion]
\label{distortion}
The distortion function $g$ is differentiable and weakly risk averse, that is, $g(x)\leq x$ for all $x\in[0,1]$.
\end{assumption}

Following \cite{quiggin93} and \cite{GhossoubHe2021}, the quantity 
$$\Delta(X) := \E[X] - DU(X)$$
is called the \textit{risk premium} associated with the random variable $X$, and weak risk aversion was shown to be equivalent to a nonnegative risk premium. Hence, an implication of weak risk aversion is that 
$$DU(L) \leq DU(\E[L]) = \E[L],$$
and we therefore obtain the following immediate result.

\begin{proposition}
\label{riskaverse g}
If $g$ satisfies Assumption \ref{distortion}, then $U(L,0)\leq -\E[L]$.
\end{proposition}

The insurer is a risk-neutral expected-utility maximizer. After receiving the premium payment $p$ from the consumer in exchange for a promised indemnity payment of $I(L)$, the end-of-period wealth of the insurer is given by 
$$p - I(L) =p - L - R(L),$$
resulting in an expected utility of 
$$p - \E[L-R(L)],$$
which is also the insurer's expected profit from the contractual agreement. 

\subsection{Hidden Risk Attitudes}\vspace{-0.25cm}

In our setting, the monopolistic insurer is able to observe the distribution of the loss random variable $L$, but not the risk attitude of the consumer. Specifically, we assume that the consumer's type $\theta$ is drawn from a continuum of types $\Theta = \left[\underline{\theta},\overline{\theta}\right] \subset \bbR_+$, and it determines the distortion function $g_\theta$ used in the consumer's utility function
$$DU_\theta (\cdot) = \int \cdot \ \mathrm{d} \, g_\theta \circ \mathbb{P}.$$

\noindent We assume that $\theta$ has a density function $f(\theta)$ with a corresponding cumulative distribution function 
$F(\theta)$ over $\Theta$, and survival function $\overline{F}(\theta) = 1 - F(\theta)$.  Hence, the utility of wealth for a consumer of type $\theta \in \Theta$ in the absence of insurance is given by
\begin{align}\label{yaari-loss-gamma} 
U_\theta(L, 0)
:=  DU_\theta(-L)
=-\int_0^{\overline{L}}\left[1-g_\theta\left(H\left(l\right)\right)\right]\mathrm{d}l.
\end{align}

Furthermore, we assume that the type space $\Theta$  is ordered, with larger values of  $\theta$  indicating a more risk-averse attitude. That is, by \cite{GhossoubHe2021}, if $\theta_1 \leq \theta_2$, then $g_{\theta_1}(x) \geq g_{\theta_2}(x)$, for all $x \in [0,1]$. Specifically, we make the following additional assumption about the collection $\left\{g_\theta\right\}_{\theta \in \Theta}$.

\begin{assumption}
\label{gtheta}
The family $\{g_\theta\}_{\theta\in\Theta}$ of distortion functions satisfies the following: 
\begin{enumerate}
\item {\bf{Continuity in Types:}}
$\{g_\theta\}_{\theta\in\Theta}$ is uniformly Lipschitz continuous in $\theta$, with common Lipschitz coefficient $c < +\infty$.
\item {\bf{Ordered Type Space:}}
$\frac{\partial g_\theta(s)}{\partial \theta} \leq 0$ for $s\in(0,1).$
\end{enumerate}
\end{assumption}

\noindent Examples that satisfy such conditions include the quadratic distortion function $g_\theta(s)=\theta s^2+(1-\theta)s$ for $\theta\in[0,1]$,  and the class  $g_\theta(s)=\frac{s}{1+\theta (1-s)}$ for $\theta\in[0,1]$, for instance.

To mitigate the potential loss, a consumer of type $\theta$  will purchase an insurance contract at a cost of $p_\theta$. The retained loss is $R_\theta(L) = L - I_\theta(L)$, where $I_\theta$ is the indemnity schedule offered by the insurer. The consumer's resulting end-of-period utility of wealth is given by 
\begin{align}
\label{Ugamma}
U_\theta\left(R_\theta,p_\theta\right)
=-p_\theta-\int_{0}^{ {\overline{L}}}\left[1-g_\theta\left(\mathbb{P}\left[R_{\theta}(L)\leq l\right]\right)\right]\mathrm{d}l.
\end{align}

It is customary in the literature to restrict the set of feasible indemnity schedules offered in the market to those functions $I:[0,\overline{L}]\rightarrow [0,\overline{L}]$ that satisfy $0 \leq I(L) \leq L$, and are such that both $I(L)$ and $R(L) = L - I(L)$ are non-decreasing functions of $L$. This restriction, often referred to as the \textit{no-sabotage condition} (e.g., \cite{CarlierDana2003b,CarlierDana2005a}) is meant to prevent \textit{ex post} moral hazard that might result from the consumer's misreporting of the true vaue of the realized loss $L$ (e.g., \cite{Huberman1983}).

Note that any indemnity function $I:[0,\overline{L}]\rightarrow [0,\overline{L}]$ that satisfies the \textit{no-sabotage} condition is a non-decreasing and $1$-Lipschitz function (e.g., \cite[Proposition 4.5]{Denneberg94}), and consequently so is the associated retention function $R$. Henceforth, we make the assumption that the market only offers indemnity functions that satisfy the \textit{no-sabotage} condition. Hence, the set of \textit{ex ante} admissible indemnity functions is given by
\begin{equation}
\label{FeasIndem}
\mathcal{I}_L:=\Big\{I: [0,\overline{L}]\rightarrow [0,\overline{L}] \ \Big\vert \ 
I(0)=0, \, 0\leq I(l_1)-I(l_2)\leq l_1-l_2, \forall\, 0\leq l_2\leq l_1\leq \overline L \Big\}.
\end{equation}

\begin{remark}
\label{CompactI}
Let $C[0,M]$ denote the set of all continuous functions on $[0,M]$ (and thus bounded), equipped with the supnorm $\Vert\cdot\Vert_{sup}$. Note that $\mathcal{I}_L$ is a uniformly bounded subset of $C[0,M]$ consisting of Lipschitz-continuous functions $[0,M] \rightarrow [0,M]$, with common Lipschitz constant $K=1$. Therefore, $\mathcal{I}_L$ is equicontinuous, and hence compact by the Arzel\`a-Ascoli Theorem \cite[Theorem IV.6.7]{DunfordSchwartz58}.
\end{remark}

Since Lipschitz-continuous functions are absolutely continuous, we can equivalently restrict the set of feasible retention functions to the set 
\begin{equation}
\label{FeasibleRetention}
\mathcal{R}_L 
:= \Big\{R: [0,\overline{L}]\rightarrow [0,\overline{L}] \ \Big\vert \ R(0)=0, 0\leq\frac{\partial R(l)}{\partial l}\leq 1, \forall\, l \in [0,\overline{L}], \text{ a.e.}\Big\}.
\end{equation}

\noindent Hence, a contract consists of a feasible indemnity function (equivalently, a feasible retention function) and an associated premium. 

\begin{definition}
A contract is a pair $(R,p) \in \mathcal{R}_L \times \bbR_+$.
\end{definition}

\subsection{Problem Formulation}\vspace{-0.25cm}

For the risk-neutral insurer, the profit from providing a contract $(R_\theta,p_\theta)$ to a consumer of type $\theta$ is
\begin{align}\label{pi}
\pi\left(R_{\theta},p_{\theta}\right)
&:=p_\theta-\mathbb{E}\left[L-R_{\theta}(L)\right]
=p_\theta-\mathbb{E}[L]+\int_0^{\overline{L}}\left[1-H\left(l\right)\right]\frac{\partial R_{\theta}(l)}{\partial l}\mathrm{d}l,
\end{align}

\noindent and the expected profit resulting from offering a menu of contracts $\left(R_{\theta}, p_{\theta}\right)_{\theta\in\Theta}$  is
\begin{align}
\label{VRT}
V\left(\left(R_{\theta}, p_{\theta}\right)_{\theta\in\Theta}\right)
:=\int_\Theta
\pi\left(R_{\theta},p_{\theta}\right)\mathrm{d}F(\theta).
\end{align}

To incentivize the consumer to participate in the market, we require \textit{individual rationality}. Additionally, to ensure that consumer reveals their true type $\theta$ and selects contracts that match their preferences, the menu of insurance contracts must also satisfy \textit{incentive compatibility}.

\begin{definition}
\label{contract}
A menu of contracts  $\left(R_{\theta},p_{\theta}\right)_{\theta\in\Theta} \in \mathcal{R}_L^\Theta \times \mathbb{R}_+^\Theta$ is said to be:
\begin{enumerate}
\item Individually rational (IR) if  $U_\theta\left(R_\theta,p_\theta\right)\geq U_\theta\left(L,0\right)$, for all $\theta\in\Theta$.
\item Incentive compatible (IC) if 
$U_\theta\left(R_\theta,p_\theta\right)\geq U_\theta\left(R_{\theta'},p_{\theta'}\right)$, for all $\theta, \theta'\in\Theta$.
\end{enumerate}  
\end{definition}

\noindent Let $\cIr$ and $\cIc$ denote the set of all IR and IC menus, and note that both sets are convex, by linearity of $U_\theta$ in \eqref{Ugamma}. Consequently $\cIr \cap \cIc$ is a convex set.

\smallskip

As the monopolist in the insurance market, the insurer aims to offer a menu that maximizes expected profit. Our goal is therefore to identify the set of all profit-maximizing IR and IC menus of contracts $(R^\ast_\theta,p^\ast_\theta)_{\theta\in\Theta}$. That is, we wish to  solve the following problem.
\begin{problem}\label{max}
\[(R^\ast_\theta,p^\ast_\theta)_{\theta\in\Theta}\in\underset{\left(R_\theta,p_\theta\right)_{\theta\in\Theta}\in\cIr\cap\cIc}\argmax \, V\left(\left(R_{\theta}, p_{\theta}\right)_{\theta\in\Theta}\right).\]
\end{problem}

\section{Characterization of IR and IC  Menus}\vspace{-0.25cm}
\label{SecIRIC}

In this section, we present preliminary results about individually rational and incentive compatible menus. These results provide an intuitive understanding of how the contract influences the insurer's expected profit and what the optimal menu could look like.

We first note an immediate implication of the definition of individual rationality.

\begin{proposition}
\label{IRmenu}
A menu of contracts  $\left(R_{\theta},p_{\theta}\right)_{\theta\in\Theta} \in \cIr$  if and only if for each $\theta \in \Theta$, $R_\theta \in \mathcal{R}_L$ and
\begin{equation}
\label{IRcondPrem}
p_\theta 
\leq 
\int_0^{\overline{L}}\left[1-g_{\theta}(H(l))\right]\left(1-\frac{\partial R_{\theta}(l)}{\partial l}\right)\mathrm{d}l.
\end{equation}
\end{proposition}

The next result provides necessary conditions for incentive compatibility. It offers a clear characterization of the premium structure and the insurer's profit, when the consumer truthfully reveals their information.

\begin{proposition}
\label{IC menu}
If $\left(R_{\theta},p_{\theta}\right)_{\theta\in\Theta}\in\cIc,$ then the following two properties hold:
\begin{enumerate}
\item For any $\theta \in \Theta$, the premium $p_\theta$ is of the form
\begin{equation}
\label{IC t}
\begin{split}
p_\theta
=p_{\underline\theta}+\int_0^{\overline{L}}\left[1-g_{\underline\theta}\left(H\left(l\right)\right)\right]\frac{\partial R_{\underline\theta}(l)}{\partial l}\mathrm{d}l
 \ - \
\int^\theta_{\underline{\theta}}\int_0^{\overline{L}} & \, \frac{\partial g_s\left(H\left(l\right)\right)}{\partial s}\frac{\partial R_{s}(l)}{\partial l}\mathrm{d}l\mathrm{d}s \\
& \quad
-\,\int_0^{\overline{L}}\left[1-g_\theta\left(H\left(l\right)\right)\right]\frac{\partial R_{\theta}(l)}{\partial l}\mathrm{d}l,
\end{split}
\end{equation} 
where $p_{\underline\theta}\in \mathbb{R}_+$ is arbitrary.

\item The insurer's expected profit is given by 
\begin{equation}
\label{IC profit}
\begin{split}
V\left(\left(R_{\theta}, p_{\theta}\right)_{\theta\in\Theta}\right)
=p_{\underline\theta}
\, + \, \int_0^{\overline{L}}\left[1-g_{\underline\theta}\left(H\left(l\right)\right)\right] & \,\frac{\partial R_{\underline\theta}(l)}{\partial l}\mathrm{d}l \\
& -  \mathbb{E}[L] 
-  \int_{\underline{\theta}}^{\overline{\theta}}\left(\int_0^{\overline{L}}J_\theta(l)\frac{\partial R_\theta(l)}{\partial l}\mathrm{d}l\right)f(\theta)\mathrm{d}\theta,
\end{split}
\end{equation}
where
\begin{align}\label{virtual value J}
J_\theta(l)
:=
H(l)-g_\theta\left(H\left(l\right)\right)+\frac{\overline{F}(\theta)}{f(\theta)}\frac{\partial g_\theta\left(H\left(l\right)\right)}{\partial \theta}.
\end{align}
\end{enumerate}
\end{proposition}

An important implication of \eqref{IC t} is that if $\left(R_{\theta},p_{\theta}\right)_{\theta\in\Theta}\in\cIc$, then for a given type $\theta \in \Theta$, the premium $p_\theta$ is fully determined by the choice of $p_{\underline{\theta}}$, the premium payment of the lowest type (the least risk averse), and the contracts $R_s$, for all types $s\leq\theta$. Such reasonable type-dependent nonlinear pricing schedules are also examined in \cite{stiglitz1977} and subsequently applied to other screening problems, as seen in \cite{chade2012}, \cite{Boonen_Zhang_2021}, and \cite{gershkov2023optimal}, for instance.  The proof of Proposition \ref{IC menu} reveals that the representation of $p_\theta$ given in \eqref{IC t} depends solely on the incentive compatibility condition. It conveys the consumers' true information—reflected through their choice of contract and corresponding utility—to the insurer. The insurer then reassesses their potential expected profit from such transactions based on this information. Further, the profit expression in \eqref{IC profit} reveals that, for any given 
$\theta$ and $l$,  the profit is a monotonic function of the marginal retention function $\frac{\partial R_\theta(l)}{\partial l}$, and  the optimal value can then be identified by analyzing the sign of
$J_\theta(l)$. A detailed analysis of this will be provided in the next section.

When consumers exhibit different attitudes toward risk, their demand for insurance varies accordingly. A highly risk-averse individual is more inclined to purchase a contract that covers a larger portion of potential random losses, thereby retaining less risk than a less risk-averse individual. For example, consider two types of individuals, $\theta_1$ and $\theta_2$, where $\theta_1<\theta_2$.  If only one type of contract, $(R_{\theta_2},p_{\theta_2})$, which provides high coverage at a high premium, is offered, individuals of type $\theta_2$ may choose to purchase it. However, individuals of type $\theta_1$
  may be more comfortable with the risk and unwilling to pay a high premium for extensive coverage. Conversely, if only $(R_{\theta_1},p_{\theta_1})$, which provides low coverage at a low premium,  is available, type $\theta_1$ 
  individuals will participate, but type $\theta_2$ individuals—who are more risk-averse and willing to pay more for greater coverage—would be underserved.
In this case, offering separate contracts tailored to each individual type improves overall welfare. If type $\theta_1$ individuals receive $(R_{\theta_1},p_{\theta_1})$
and type $\theta_2$ individuals receive $(R_{\theta_2},p_{\theta_2})$, this design ensures that each consumer benefits. When the coverage level of a menu of contracts $(R_{\theta},p_\theta)_{\theta \in \Theta}$ follows such a monotonic structure for any level of risk, meaning that more risk-averse individuals receive greater coverage, thereby retaining less loss, we say that the $\{R_{\theta}\}_{\theta \in \Theta}$ is \textit{submodular}. This concept is formally defined as follows.

\begin{definition}\label{submodular}
A collection $\{R_{\theta}\}_{\theta \in \Theta}$ of retention functions is submodular if 
$\frac{\partial R_\theta(l)}{\partial l}$
is non-increasing in $\theta$  for all $ l\in[0,\overline{L}].$
\end{definition}
 
Combined with Proposition \ref{IC menu}, the submodularity of the retention function allows us to derive a necessary and sufficient condition for a menu to be IC, as shown in the following result. 
 
\begin{proposition}
\label{submodular IC}
If $\{R_\theta\}_{\theta\in\Theta}$ is submodular, then the menu $\left(R_{\theta},p_{\theta}\right)_{\theta\in\Theta}\in\cIc$ if and only if $\{p_\theta\}_{\theta\in\Theta}$ satisfies \eqref{IC t}.
\end{proposition}

The necessary part of this corollary is obvious from Proposition \ref{IC menu}. The sufficient part holds when $R$
is monotone with the risk type. This corollary indicates that an IC menu is equivalent to any insurance menu with monotonic insurance coverage and premium  determined by \eqref{IC t}. This aligns with the condition for a submodular retention function to be an IC menu when the hidden information is the loss distribution ratrher than the consumer's level of risk aversion. This is characterized in \citet[Proposition 1(ii)]{gershkov2023optimal}. Notably, in \cite{gershkov2023optimal}, submodular retention implies offering greater insurance coverage to individuals characterized by a stochastically larger loss distribution. 

In addition to the IC condition, the menus of contracts must be designed to ensure that consumers are willing to participate. Thus, we also require that the menus of contracts be IR. The following result shows that an IC menu is IR if and only if the contract offered to the lowest type is IR.  

\begin{proposition}
\label{IR lowest}
If $\left(R_{\theta},p_{\theta}\right)_{\theta\in\Theta}\in\cIc$, then $\left(R_{\theta},p_{\theta}\right)_{\theta\in\Theta}\in \cIr$ if and only if $(R_{\underline\theta},p_{\underline\theta})\in\cIr.$ 
\end{proposition}

\noindent This aligns with the situation where private information is the loss distribution rather than the level of risk aversion, as shown in Lemma 1 of \cite{gershkov2023optimal}. Based on Proposition \ref{IR lowest}, we can determine an interval of values for $p_{\underline\theta}$  that ensures that an IC menu is IR. This is given below.
  
\begin{corollary}
\label{IC t interval}
If $\{R_\theta\}_{\theta\in\Theta}$ is submodular, then $\left(R_{\theta},p_{\theta}\right)_{\theta\in\Theta}\in \cIr\cap\cIc$ if and only if $\{p_\theta\}_{\theta\in\Theta}$ satisfies \eqref{IC t}, with
\[p_{\underline\theta}\leq \int_0^{\overline{L}}\left[1-g_{\underline\theta}(H(l))\right]\left(1-\frac{\partial R_{\underline\theta}(l)}{\partial l}\right)\mathrm{d}l.\]
\end{corollary}

\noindent This result follows directly from Proposition \ref{submodular IC} and Proposition \ref{IR lowest}. It implies that whenever we have an IC menu with monotone insurance coverage, it is enough to verify the premium of the lowest risk type to determine whether the menu is individually rational.

In the next section, we will use the results given above in order to characterize the profit-maximizing, IR, and IC menus.


\section{Solution and Properties of the Optimal Menu}\vspace{-0.25cm}
\label{Profit-O-menu}

\subsection{Equilibrium Contracts with Full Information}\vspace{-0.25cm}

As argued by \cite{stiglitz1977}, in a monopoly market with a risk-neutral insurer and a risk-averse EU-maximizing consumer, and under perfect information, equilibria consist of full-insurance contracts at premia that make the consumer indifferent between purchasing insurance and foregoing coverage. That is, all the consumer surplus involved in the reduction of risk is extracted by the monopoly, as is classically understood. In this section we show that this insight still holds when the consumer is Yaari DU-maximizer.

To that end, consider the benchmark case of full information in the market, whereby the risk aversion of the consumer is observable by the insurer. This is tantamount to assuming that the type space is a singleton of the form $\Theta =\{\theta_0\}$. In this case, the insurer no longer needs to offer menus of contracts, but rather a single contract to the consumer, which is \textit{de facto} incentive compatible. The monopolist's problem is therefore to design a contract that is profit-maximizing and individually rational:

\begin{problem}
\label{max2}
$$\underset{(R,p) \in \mathcal{R}_L \times \bbR_+} \sup \, \Big\{\pi(R,p): U_{\theta_0}(R,p) \geq U_{\theta_0}(L,0)\Big\}.$$
\end{problem}

Noting that for all $R \in \mathcal{R}_L$,
$$\int_{0}^{ {\overline{L}}}\left[1-g_{\theta_0}\left(\mathbb{P}\left[R(L)\leq l\right]\right)\right] \, \mathrm{d}l
=
\int_{0}^{ {\overline{L}}}\left[1-g_{\theta_0}\left(H(l)\right)\right] \, R^\prime(l)\ \mathrm{d}l,$$

\noindent and using \eqref{Ugamma} and \eqref{pi}, Problem \ref{max2} can be rewritten as:
\begin{equation}
\label{max2b}
\underset{(R,p) \in \mathcal{R}_L \times \bbR_+} \sup \, 
\Bigg\{
p-\mathbb{E}[L]+\int_0^{\overline{L}}\left[1-H\left(l\right)\right] \, R^\prime(l)\,\mathrm{d}l \ \Big\vert \  
p
\leq 
\int_{0}^{ {\overline{L}}}\left[1-g_{\theta_0}\left(H(l)\right)\right] \, \left[1-R^\prime(l)\right]\, \mathrm{d}l
\Bigg\}.
\end{equation}

The following result shows that in the case of perfect information in our monopoly market, full insurance is optimal, at a premium that makes the consumer indifferent between insurance and no insurance.

\begin{proposition}
\label{one}
The optimal solution to Problem \ref{max2} is given by the contract
$$(R^*_{\theta_0}, p^*_{\theta_0}) = \left(0,\displaystyle\int_{0}^{\overline{L}}\left[1-g_{\theta_0}\left(H\left(l\right)\right)\right]\mathrm{d}l\right).$$

\noindent Moreover, 
$$U_{\theta_0}(R^*_{\theta_0}, p^*_{\theta_0})
=U_{\theta_0}(L,0).
$$
\end{proposition}

The optimal contract entails zero retention, that is, full insurance, at a premium of 
$$p^*_{\theta_0} = \displaystyle\int_{0}^{\overline{L}}\left[1-g_{\theta_0}\left(H\left(l\right)\right)\right]\mathrm{d}l = -U_{\theta_0}(L,0) = -DU(-L),$$
which leads to all the consumer surplus being absorbed by the monopoly.

\subsection{Equilibrium Contracts with Hidden Risk Attitudes}\vspace{-0.25cm}

We now extend the benchmark case of a market with full information to a setting with imperfect information, in which the consumer's risk aversion is unobservable to the insurer. Theorem \ref{profit solution} characterizes the profit-maximizing menu under imperfect information, ensuring incentive compatibility  and individual rationality, with the solution expressed in terms of marginal loss retention.

\begin{theorem}
\label{profit solution}
Suppose that $J_\theta(l)$ given in \eqref{virtual value J}
is non-decreasing in $\theta$, for all $l$. An optimal solution $\left(R^\ast_\theta,p^\ast_\theta\right)_{\theta\in\Theta}$ for  Problem \ref{max}
is characterized by the following:

\begin{enumerate}
\item For each $\theta\in\Theta$, the retention function $R^*_\theta$ satisfies
\begin{equation}\label{band-band}
\frac{\partial R^\ast_\theta(l)}{\partial l}=
\left\{
\begin{array}[c]{ll}%
0, &  J_\theta(l)>0,\vspace{0.2cm}\\
\in[0,1], &  J_\theta(l)=0,\vspace{0.2cm}\\
1, & J_\theta(l)< 0.
\end{array}
\right.  
\end{equation}

\item For each $\theta\in\Theta$, the corresponding premium $p^\ast_\theta$ is given by:
\begin{equation}
\label{optimal t} 
p^\ast_\theta
:=\int_0^{\overline{L}}\left[1-g_{\underline\theta}(H(l))\right]\mathrm{d}l
- \int^\theta_{\underline{\theta}}\int_0^{\overline{L}} \,\frac{\partial g_s\left(H\left(l\right)\right)}{\partial s}\,\frac{\partial R^\ast_{s}(l)}{\partial l}\mathrm{d}l\mathrm{d}s -\int_0^{\overline{L}}\left[1-g_\theta\left(H\left(l\right)\right)\right]\frac{\partial R^\ast_{\theta}(l)}{\partial l}\,\mathrm{d}l.
\end{equation} 
\end{enumerate}

\noindent Moreover, the collection $\{R^*_{\theta}\}_{\theta \in \Theta}$ of optimal retention functions is submodular.
\end{theorem}

\noindent The above results asserts that, under the assumption that, for each $l \in [0,\overline{L
}]$, the function $\theta \mapsto J_\theta(l)$ is non-decreasing, the optimal menu of contracts consists of a collection of layered indemnity schedules, and hence layered corresponding retention functions. Specifically, the marginal retention at a loss of value $l \in [0,\overline L]$ is $0$ when $J_\theta(l)>0$, corresponding to full insurance. When $J_\theta(l)<0$, the marginal retention is $1$, corresponding to no insurance for that level of loss. When $J_\theta(l)=0$, the optimal retention allows for some flexibility, as long as feasibility is maintained. Additionally, at an optimum, the premium determined by a non-linear function of the insurance coverage as described in \eqref{optimal t}. In other words, a profit-maximizing, IR, and IC menu of contracts leads to a separating equilibrium. We comment further below on the monotonicity of $J_\theta(l)$ in $\theta$.

Additionally, the optimal menu has an intuitive property. Since the collection $\{R^*_{\theta}\}_{\theta \in \Theta}$ of optimal retention functions is submodular, it follows at any loss level $l \in [0,\overline L]$, the marginal loss retention of a lower risk type always exceeds that of a higher risk type. A less risk-consumer  retains more marginal loss than a higher risk-averse consumer at an optimum.

While in competitive insurance markets only differences in risk distribution affect equilibria and the nature of separating contracts, in a monopoly both risk and risk attitude have an impact on equilibrium contracts, as observed by \cite{stiglitz1977}. The latter showed if the loss distribution is known by the monopolist insurer, but the consumer's level of risk aversion is hidden from the insurer, then equilibrium contracts can have different forms, and they can be separating or pooling equilibria. We show below that this is not the case in our setting, and that separating equilibria always outperform pooling equilibria.

\begin{proposition}
\label{ComparisionPooling}
Suppose that $J_\theta(l)$ given in \eqref{virtual value J}
is non-decreasing in $\theta$ for all $l$.      The separating equilibrium described in Theorem \ref{profit solution} is more valuable to the insurer than a pooling equilibrium contract.
\end{proposition}

\noindent The above result demonstrates that in a monopolistic market, selecting contracts that satisfy the IR and IC conditions, as outlined in Theorem \ref{profit solution}, always outperforms offering a single contract. This result extends a key finding of \cite{Boonen_Zhang_2021}. More properties of the optimal menu will be examined in Section \ref{sec: prop opt menu}.

\subsection{The Function $\theta \mapsto J_\theta(l)$}\vspace{-0.25cm}

\subsubsection{Two Components}\vspace{-0.25cm}

Next, we provide an analysis of how the virtual value function $$J_\theta(l)
=
H(l)-g_\theta\left(H\left(l\right)\right)
+
\frac{\overline{F}(\theta)}{f(\theta)}\frac{\partial g_\theta\left(H\left(l\right)\right)}{\partial \theta}$$ 
affects the insurer's profit, and we provide conditions along with interpretations for the monotonicity of the map $\theta \mapsto J_\theta(l)$ to hold.

First, note that $J_\theta(l)$ can be decomposed into two components. The first component is
$$H(l)-g_\theta\left(H\left(l\right)\right),$$
which is nonnegative by Assumption \ref{distortion} (weak risk aversion), and non-decreasing in $\theta$ by Assumption \ref{gtheta} (ordered type space). This quantity can be seen as a proxy for the level of risk aversion of type $\theta$ at the point $H(l) \in [0,1]$, since weak risk aversion of type $\theta$ is equivalent to $g_\theta(x) \leq x$, for $x\in [0,1]$. 

The second component of $J_\theta(l)$ is
$$\frac{\overline{F}(\theta)}{f(\theta)}\,\frac{\partial g_\theta\left(H\left(l\right)\right)}{\partial \theta},$$
which can be interpreted as the information rent. It serves as a cost since $\frac{\partial g_\theta\left(H\left(l\right)\right)}{\partial \theta}$ is non-positive, by Assumption \ref{gtheta}. This arises because consumers have private knowledge about their risk attitudes, which the insurer does not observe directly. The insurer must account for this private information when designing contracts to mitigate adverse selection. If this component is increasing in $\theta$,  then marginal changes in coverage or retention result in a smaller reduction in the insurer's profit for higher values of $\theta$. This implies that as risk aversion increases, it becomes less costly to design incentive compatible contracts for more risk averse individuals.

Together, these two components determine how the insurer's profit changes for any marginal increase in coverage or marginal decrease in retention.

\subsubsection{Monotonicity}\label{monotonicity}\vspace{-0.25cm}

Next, we provide some intuitive sufficient conditions for the monotonicity of the map $\theta \mapsto J_\theta(l)$, for a given $l \in [0,\overline L]$. First, as mentioned above, the first component is  non-decreasing in $\theta$, by assumption. For the second component, differentiating with respect to $\theta$ yields
\begin{align}
\label{J derivative}
\frac{\partial }{\partial \theta}\left(\frac{\overline{F}(\theta)}{f(\theta)}\,\frac{\partial g_\theta\left(H\left(l\right)\right)}{\partial \theta}\right)
&=\left(\frac{\overline{F}(\theta)}{f(\theta)}\right)^\prime\frac{\partial g_\theta\left(H\left(l\right)\right)}{\partial \theta}+\frac{\overline{F}(\theta)}{f(\theta)}\frac{\partial^2 g_\theta\left(H\left(l\right)\right)}{\partial \theta^2}\nonumber\\
&=-\frac{\partial g_\theta\left(H\left(l\right)\right)}{\partial \theta}\left(-\frac{\overline{F}(\theta)}{f(\theta)}\right)^\prime\
+
\frac{\overline{F}(\theta)}{f(\theta)}\,\frac{\partial^2 g_\theta\left(H\left(l\right)\right)}{\partial \theta^2}.
\end{align}

\noindent This quantity is non-negative if the following two conditions hold jointly:
\begin{enumerate}
    \item The hazard rate $\nicefrac{f}{\overline{F}}$ is non-decreasing over $\Theta;$
    \item For each $s\in(0,1)$, the function $\theta \mapsto g_\theta(s)$ is convex.
\end{enumerate}

The first condition is imposed on the distribution $F$ of types over the type space $\Theta$. A non-decreasing hazard rate for the distribution $F$ ensures that the first term in \eqref{J derivative} is non-negative, since $\frac{\partial g_\theta\left(H\left(l\right)\right)}{\partial \theta}\leq 0$, by assumption.

The second condition implies that the second term in \eqref{J derivative} is non-negative. Together with Assumption \ref{gtheta}, this condition implies that the utility specified in \eqref{yaari-loss-gamma} is non-increasing and convex with respect to the risk aversion parameter $\theta$. In other words, the difference in the utility of wealth between risk types diminishes as the type increases. This observation is consistent with commonly used utility functions, such as exponential utility functions or power utility functions, for instance.

Many distortion functions satisfy the above two conditions. For example:  $g_\theta(s)=\theta s^2+(1-\theta)s,$ where $\theta\in[0,1]$;  $g_\theta(s)=\frac{s}{1+\theta (1-s)},$ where $\theta\in[0,1]$; and the family of exponential distortions: $g_\theta(s)=\frac{e^{-\theta(1-s)}-e^{-\theta}}{1-e^{-\theta}},$ where $\theta\in[0,1].$

If the above two conditions hold, then $J_\theta(l)$ is a monotonic function of $\theta$ for any $l\in[0,\overline{L}]$.  A similar monotonicity condition is also introduced when the hidden information pertains to the risk distribution, as discussed in \citet{gershkov2023optimal}. In their study, a similar function to 
$J$ appears. To ensure monotonicity of this function, the authors assume that the type distribution has a non-decreasing hazard rate. Additionally, they impose the following other conditions:
\begin{enumerate}
\item The distortion function in Yaari's dual utility framework must be convex. This ensures that the utility function reflects aversion to mean-preserving spreads (strong risk aversion).
\item The collection of loss distributions indexed by types must exhibit stochastic concavity with respect to the type parameter. This implies that as the risk type increases, the risk distributions satisfy a concave ordering.  
\item For the highest risk type $\overline{\theta}$, with cumulative distribution function $ H_{\overline{\theta}} $, it is assumed that $ g(H_{\overline{\theta}}(0)) \geq 1 $. This condition corresponds to the requirement that the probability of an accident for the highest risk type must meet a minimum threshold.
\end{enumerate}

\subsection{Properties of Optimal Menus}
\label{sec: prop opt menu}
\vspace{-0.25cm}

The following proposition elaborates on several properties of the optimal menu.

\begin{proposition}
\label{menu properties} Suppose that $J_\theta(l)$ given in \eqref{virtual value J}
is non-decreasing in $\theta$, for all $l$.
The optimal menu $\left(R^\ast_\theta,p^\ast_\theta\right)_{\theta\in\Theta}$ in Theorem \ref{profit solution} satisfies the following properties.
\begin{enumerate}
\item The retention function $R^\ast_\theta$ decreases with $\theta$, and the premium $p^\ast_\theta$ increases with $\theta$. 

\item For the most risk averse consumer, $R_{\overline{\theta}}(l)=0$ for all $l\in[0,\overline{L}]$.

\item For the least risk averse consumer, $U_{\underline\theta}(R^\ast_{\underline\theta},p^\ast_{\underline\theta})=U_{\underline{\theta}}(L,0).$ 

\item The consumer's utility $U_\theta\left(R^\ast_{\theta},p^\ast_{\theta}\right)$ decreases with $\theta$. Moreover, it is convex in $\theta$ if $g_\theta$ is convex in $\theta$. 

\item The insurer's expected profit from a type-$\theta$ contract, that is $\pi\left(R^\ast_{\theta},p^\ast_{\theta}\right)$, increases with $\theta$.

\item The insurer's overall expected profit is positive: $V\left(\left(R^\ast_{\theta}, p^\ast_{\theta}\right)_{\theta\in\Theta}\right) > 0.$
\end{enumerate}
\end{proposition}

This proposition outlines key properties of the profit-maximizing menu. The first result shows that insurance coverage and premia are monotone in the level of risk aversion. A more risk averse consumer is offered a contract with greater coverage but at a higher price. This finding aligns with \citet[Theorem 1]{gershkov2023optimal} and \cite{chade2012}, in a setting with the hidden information is the loss distribution rather than the risk attitude of the consumer. However, the key distinction is that in \cite{chade2012}, the monotonicity of coverage and premia is imposed by assumption. Indeed, their approach relies on the assumption that the consumer's utility satisfies the single-crossing property, which implies that the second derivative of the utility with respect to the retention and the premium is positive, ensuring the monotonicity of coverage and premia with risk types, as the authors show. In contrast, our results establish this monotonicity as a property of optimal menus. Rather than imposing conditions directly on the consumer's utility, we impose a monotonicity condition on the virtual value function $J_\theta(l)$, as is the case in \cite{gershkov2023optimal}.

The second result demonstrates that the most risk-averse consumer will be offered full insurance coverage. This phenomenon, referred to as \textit{efficiency at the top} in \cite{chade2012}, also holds when the hidden information pertains to the loss distribution rather than the consumer's risk attitude, as shown in \citet[Theorem 1]{gershkov2023optimal}. 

The third result shows that the monopoly will absorb all surplus from the least risk-averse consumer, and the contract offered to this consumer leaves them indifferent between participation and non-participation. This property is also observed in \citet[Theorem 1(iv)]{chade2012} and is particularly evident in \cite{Boonen_Zhang_2021}, where the insurance market consists of only two types of consumers with hidden risk attitude.

The fourth result elicits an interesting monotonicity property of the consumer's utility of wealth at an optimum, as a function of their level of risk aversion, i.e., their risk type. As discussed earlier, if $g_\theta$ is convex in $\theta$, then the utility of wealth of a type-$\theta$ consumer in the absence of insurance (i.e., $U_{\theta}(L,0)$) is decreasing and convex in $\theta$. After trading, the consumer's utility becomes $U_{\theta}(R_\theta,p_\theta)$, which decreasing and convex in $\theta$, but attains a higher value than $U_{\theta}(L,0)$, with equality holding for the type $\underline{\theta}$ consumer. This echoes the finding of \citet[Theorem 1]{gershkov2023optimal}.  In \citet[Lemma 4]{chade2012}, it is shown that the utility is a decreasing function of the risk type, but the convexity is not discussed.

The fifth result demonstrates that consumers with a higher level of risk aversion, who are willing to pay more for a product with a higher coverage, are generally more valuable to the insurer, in that they induce a higher expected profit for the insurer. This contradicts the findings in scenarios with a hidden loss distribution, such as in \cite{chade2012} and \cite{gershkov2023optimal}, where the most profitable consumers are always of intermediate, rather than extreme, types.

The final result asserts that the insurer is expected to earn a positive profit, making it profitable for the monopoly to design contracts even in the presence of asymmetric information. This finding aligns with established results on optimal insurance under adverse selection in a monopoly market, as in \citet[Theorem 1(vi)]{chade2012} and \citet[Lemma 2]{gershkov2023optimal}.

\begin{remark}
It is worth noting that in both \citet[Proposition 2]{gershkov2023optimal} and \citet[Theorem 2 (ii)]{chade2012}, where the loss distribution is hidden from the insurer, the authors show that with an optimal menu that fully sorts consumers, the insurer's expected profit increases as consumers become more risk averse. However, in our setting, even with the optimal menu described in Theorem \ref{profit solution}, the insurer's profit does not necessarily increase when the loss distribution shifts under first-order stochastic dominance, whether increasing or decreasing. Indeed, the profit with $(R_\theta^\ast, p_\theta^\ast)$ can be rewritten as
\begin{align*}
&\pi(R^\ast_\theta,p^\ast_\theta)
=\int_0^{\overline{L}}\left[1-g_{\underline{\theta}}\left(H\left(l\right)\right)\right]\mathrm{d}l-\int_0^{\overline{L}}\left(\int^\theta_{\underline{\theta}}\frac{\partial R^\ast_{s}(l)}{\partial l}\mathrm{d}g_s\left(H\left(l\right)\right)\right)\mathrm{d}l-\int_0^{\overline{L}}\left[1-g_{{\theta}}\left(H\left(l\right)\right)\right]\frac{\partial R^\ast_\theta(l)}{\partial l}\mathrm{d}l\\
&\quad\quad\quad\quad\quad\quad-
\int_0^{\overline{L}}\left[1-H\left(l\right)\right]\left(1-\frac{\partial R^\ast_\theta(l)}{\partial l}\right)\mathrm{d}l\\
&=\int_0^{\overline{L}}\left[1-g_{\underline{\theta}}\left(H\left(l\right)\right)\right]\mathrm{d}l-\int_0^{\overline{L}}\left[1-g_{{\theta}}\left(H\left(l\right)\right)\right]\frac{\partial R^\ast_\theta(l)}{\partial l}\mathrm{d}l-
\int_0^{\overline{L}}\left[1-H\left(l\right)\right]\left(1-\frac{\partial R^\ast_\theta(l)}{\partial l}\right)\mathrm{d}l
\\
&\quad\quad\quad\quad\quad\quad-\int_0^{\overline{L}}\left(\frac{\partial R^\ast_{\theta}(l)}{\partial l}g_\theta\left(H\left(l\right)\right)-\frac{\partial R^\ast_{\underline{\theta}}(l)}{\partial l}g_{\underline{\theta}}\left(H\left(l\right)\right)-\int_{\underline{\theta}}^\theta g_s\left(H\left(l\right)\right)\frac{\partial}{\partial s}\frac{\partial R^\ast_s(l)}{\partial l}\mathrm{d}s\right)\mathrm{d}l\\
&=-\int_0^{\overline{L}}\left[1-H(l)\right]\left(1-\frac{\partial R^\ast_\theta(l)}{\partial l}\right)\mathrm{d}l+\int_0^{\overline{L}}\left[1-g_{\underline{\theta}}\left(H\left(l\right)\right)\right]\left(1-\frac{\partial R^\ast_{\underline\theta}(l)}{\partial l}\right)\mathrm{d}l
+\int_0^{\overline{L}}\left(\frac{\partial R^\ast_{\underline\theta}(l)}{\partial l}-\frac{\partial R^\ast_\theta(l)}{\partial l}\right)\mathrm{d}l\\
&\quad\quad\quad\quad\quad\quad+\int_0^{\overline{L}}\int_{\underline{\theta}}^\theta g_s\left(H\left(l\right)\right)\frac{\partial}{\partial s}\frac{\partial R^\ast_s(l)}{\partial l}\mathrm{d}s\mathrm{d}l.
\end{align*}

\noindent The first term, $\displaystyle\int_0^{\overline{L}}\left[1-H(l)\right]\left(1-\frac{\partial R^\ast_\theta(l)}{\partial l}\right)\mathrm{d}l,$ represents the expected coverage of the type-$\theta$ consumer. This term increases as $H$ increases in the sense of first-order stochastic dominance, which subsequently leads to a decrease in the insurer's profit. The remaining three items represent the premium charged to the consumers, and the result shows that the endogenous premium $p_\theta$ increases when $H$ becomes more risky, leading to an increase in the insurer's profit. Combining these two effects, one cannot determine the direction of change in the insurer's total profit for this type of consumer, nor can one conclusively determine the resulting expected profit.
\end{remark}

\subsection{Special Cases}\vspace{-0.25cm}

Another implication of Theorem \ref{profit solution} is that when $J_\theta(l)$ is monotone in $\theta,$ the contract menu consists of layered options, with any additional marginal loss either fully borne by the consumer or entirely transferred to the insurer. We now examine two special cases where the layered contracts are among the most popular types of insurance contracts: the deductible contract menu and the coverage limit contract.

\begin{theorem}\label{special case}
Suppose that $J_\theta(l)$, as defined in \eqref{virtual value J}, is non-decreasing with respect to $\theta$, for all $l.$
\begin{enumerate}
\item If for each $\theta<\overline{\theta}$ there exists a unique $l_\theta\in [0,\overline{L}]$ such that $J_\theta(l)\leq 0$ for $l\leq l_\theta$ and $J_\theta(l)\geq0$ for $l\geq l_\theta,$ then the optimal contract is of the deductible type, with the menu given by $\left(\max(L,l_{\theta}), p_{\theta}\right)_{\theta\in\Theta}.$

\item If for each $\theta<\overline{\theta}$ there exists a unique $l_\theta\in[0,\overline{L}]$ such that $J_\theta(l)\geq 0$ for $l\leq l_\theta$ and $J_\theta(l)\leq0$ for $l\geq l_\theta,$ then the optimal contract is of coverage limit type, with the menu given by $\left((L-l_{\theta})_+, p_{\theta}\right)_{\theta\in\Theta}.$
\end{enumerate}
\end{theorem}

The above results imply that if $J_\theta(l)$ crosses the zero line from below for all $\theta$, then a deductible menu will be offered to the consumers. Furthermore, since  Theorem \ref{profit solution} establishes that the solution is submodular, the deductible level $l_\theta$ decreases as $\theta$ increases.  Conversely, if $J_\theta(l)$ crosses the zero line from above for all $\theta$, then a coverage limit contract is provided, and the limit $l_\theta$ increases with $\theta.$ The following are two examples that illustrate the optimality of deductible contracts and coverage limit contracts.

\begin{example}[Optimality of a deductible menu]
\label{deductible menu}
Suppose that each $g_\theta$ belongs to the family of inverse S-shaped distortion (ISSD) functions introduced in \cite{xu2013a}:
\begin{equation*}
g_\theta(s)=
\left\{
\begin{array}[c]{ll}%
(1-\theta)s-(1-\theta)s^2, & \textit{if} \ 0\leq s\leq \frac{1}{2},\vspace{0.2cm}\\
(3+\theta)s^2-(3+\theta)s+1, & \textit{if} \ \frac{1}{2}\leq s\leq 1.
\end{array}
\right.  
\end{equation*}

\noindent Suppose that $\theta$ follows a uniform distribution on $[0,1]$.  Then  
\begin{align*}
J_\theta(l)&=H(l)-g_\theta\left(H\left(l\right)\right)+(1-\theta)\frac{\partial g_\theta\left(H\left(l\right)\right)}{\partial \theta}\vspace{0.4cm}\\
&=\left\{
\begin{array}[c]{ll}%
\theta H\left(l\right)+(1-\theta)H^2\left(l\right)+(1-\theta)\left(H^2\left(l\right)-H\left(l\right)\right),  & \textit{if} \ 0\leq H\left(l\right)\leq \frac{1}{2},\vspace{0.4cm}\\
(4+\theta)H\left(l\right)-(3+\theta)H^2\left(l\right)-1+(1-\theta)\left(H^2\left(l\right)-H\left(l\right)\right), & \textit{if} \ \frac{1}{2}\leq H\left(l\right)\leq 1.\vspace{0.2cm}
\end{array}
\right.  \vspace{0.9cm}\\
&=\left\{
\begin{array}[c]{ll}%
H\left(l\right)\left((2-2\theta)H\left(l\right)+(2\theta-1)\right),  & \textit{if} \ 0\leq H\left(l\right)\leq \frac{1}{2},\vspace{0.4cm}\\
\left(-2-2\theta\right)H^2\left(l\right)+\left(3+2\theta\right)H\left(l\right)-1, & \textit{if} \ \frac{1}{2}\leq H\left(l\right)\leq 1.
\end{array}
\right.  
\end{align*}

\noindent For a given $\theta\in\left[0,\frac{1}{2}\right),$ it is evident that $J_\theta(0)=J_\theta(\overline{L})=0.$ When  $l\in\left(0,H^{-1}\left(\frac{1}{2}\right)\right]$, $\left(2-2\theta\right)H\left(l\right)+(2\theta-1)$
is an increasing function of $l$ for a given $\theta$, and it crosses zero exactly once. Denote the root by $l_\theta$. For $l\in\left( H^{-1}\left(\frac{1}{2}\right),\overline{L}\right),$ the function $\left(-2-2\theta\right)H^2\left(l\right)+\left(3+2\theta\right)H\left(l\right)-1$ is a downward-opening quadratic function of $H(l).$ This quadratic function has two roots, $H^{-1}\left(\frac{1}{2\theta+2}\right)$ and $H^{-1}(1)$, but neither lies within the interval. Thus, there is no root in this range. Consequently, $J_\theta(l)\leq0$ when $l<l_\theta,$ and $J_\theta(l)\geq0$ when $l\geq l_\theta.$ By Theorem \ref{special case}, the optimal contract is a deductible, with the deductible level given by
\[l_{\theta}=H^{-1}\left(\frac{1-2\theta}{2-2\theta}\right).\]
If $\theta\geq\frac{1}{2},$ $J_\theta(l)$ remains non-negative for all $l,$ implying that $l_\theta=0.$  In this case, full insurance is optimal.
\end{example}

The above example examines the scenario where the consumer of type $\theta$ uses a ISSD function $g_\theta$, and is therefore mostly concerned with the tails of the loss distribution, that is relatively small and relatively large losses. The results show that for consumers who are less risk averse $(\theta<\frac{1}{2})$, partial insurance will be provided in the form of a deductible contract. For consumers who are more risk averse $(\theta\geq\frac{1}{2}),$ full insurance is provided under each contract, leading to a form of pooling at the top (pooling with full insurance).

\begin{example}[Optimality of a coverage limit menu]
\label{coverage limit menu}
Suppose that  $g_\theta(s)=\frac{s}{1+\theta (1-s)}$, where $\theta\in[0,1]$. Suppose further that $\theta$ follows a non-decreasing hazard rate distribution. We obtain that
\begin{align*}
J_\theta(l)
&=H(l)-g_\theta\left(H\left(l\right)\right)+\frac{\overline{F}(\theta)}{f(\theta)}\frac{\partial g_\theta\left(H\left(l\right)\right)}{\partial \theta}\\
&=\frac{\theta H(l)\left(1-H(l)\right)}{1+\theta\left(1-H(l)\right)}-\frac{\overline{F}(\theta)}{f(\theta)}\frac{H(l)\left(1-H(l)\right)}{\left(1+\theta\left(1-H(l)\right)\right)^2}\\
&=\left(\theta-\frac{\overline{F}(\theta)}{f(\theta)}\frac{1}{1+\theta\left(1-H(l)\right)}\right)\frac{H\left(l\right)-H^2\left(l\right)}{1+\theta\left(1-H(l)\right)}.
\end{align*}

\noindent For a fixed $\theta,$ $J_\theta(l)$ always crosses zero from above since the function $\frac{H\left(l\right)-H^2\left(l\right)}{1+\theta\left(1-H(l)\right)}$ remains non-negative on $[0,\overline{L}],$ and $\theta-\frac{\overline{F}(\theta)}{f(\theta)}\frac{1}{1+\theta\left(1-H(l)\right)}$ is  a decreasing function of $l$. This implies that the optimal menu is a coverage limit contract by Theorem \ref{special case}.  If $\theta<\frac{1}{1+\theta}\frac{\overline{F}(\theta)}{f(\theta)},$ then $J_\theta(l)\leq0,$ and full retention (no coverage) is optimal, with $l_\theta=0.$ If $\frac{1}{1+\theta}\frac{\overline{F}(\theta)}{f(\theta)}\leq\theta<\frac{\overline{F}(\theta)}{f(\theta)}$, then the coverage limit is given by
\[l_{\theta}=H^{-1}\left(1+\frac{1}{\theta}-\frac{\overline{F}(\theta)}{\theta^2 f(\theta)}\right).\]
If $\theta>\frac{\overline{F}(\theta)}{f(\theta)},$  $J_\theta(l)\geq0$  and thus $l_\theta=\overline{L}$, or full insurance is optimal.
\end{example}

If the consumer places greater weight on higher levels of loss, as illustrated in Example \ref{coverage limit menu} with a convex distortion, the coverage limit contract becomes optimal. The results indicate that no insurance is provided to the least risk-averse consumer, full insurance is provided to the most risk-averse consumer, and partial insurance is offered to those in between. This example illustrates that pooling occurs at both the bottom (pooling with no insurance) and the top  when the consumer's risk attitude is hidden information. In this case, coverage denial may  occur, and typically, only a few contracts are offered, most of which providing full insurance.

These two types of contracts are also analyzed in \citet[Theorem 3]{gershkov2023optimal}. The authors argue that, for any given expected indemnification, a deductible contract is the most preferable option for a highly risk-averse consumer, whereas a coverage limit contract is favored when it must satisfy the IC condition. In \citet[Example 2]{gershkov2023optimal}, where hidden information arises from the probability of a loss, the deductible menu is selected, with pooling observed both at the bottom  and at the top. In \citet[Example 3]{gershkov2023optimal}, where hidden information pertains to the loss size, the coverage limit menu emerges as the optimal choice.

To conclude this section, we have shown that in a monopoly market without hidden risk attitudes, full insurance maximizes the insurer’s profit, with the premium set to bind the consumer’s welfare constraint. However, when the consumer's risk attitude is hidden from the insurer, the profit-maximizing, IR, and IC menu of contracts consists of a set of layered contracts. This menu yields a strictly positive profit for the monopoly, and, under these contracts, less risk-averse consumers bear a larger portion of the incurred loss at a relatively lower price, while more risk-averse consumers retain a smaller portion of the loss and pay a higher premium.

\section{The Effect of Friction Costs on Equilibria}\vspace{-0.25cm}
\label{SecCost}

Thus far, the monopoly market that we consider did not account for any friction costs incurred by the insurer. The cost of insurance was simply the expected value of the indemnity payment, and the revenue to the insurer was the insurance premium collected from the consumer. As a result, the insurer's profit was the premium payment less the expected indemnity payment. 

In reality, insurers often incur additional ``friction costs'', such as claims processing costs, administrative expenses, actuarial loading, etc. Moreover, those costs can have a significant impact on the shape of contracts offered. This was shown, for instance, by \cite{RAVIV1979},  \cite{Huberman1983}, and \cite{CarlierDana2003b}, in a setting of perfect information.

\subsection{Equilibria with Costs and Hidden Risk Attitudes}\vspace{-0.25cm}

In this section, we extend our model to account for such costs. Specifically, we consider the case of a fixed cost that occurs only when insurance is provided. We use the same cost function as the one proposed in \cite{CHADE2020105085}, which incorporates a fixed friction cost in addition to the  indemnification cost. This friction cost arises whenever insurance is provided, i.e., $R_{\theta}(l)\not\equiv l$. Let the friction cost be denoted by $k>0$. Notably, this cost does not alter any fundamental properties of the IR or IC menu; rather, it only impacts the insurer's profit to some extent. Consequently, Proposition \ref{IC menu}(1), Proposition \ref{submodular IC}, Proposition \ref{IR lowest}, and Corollary \ref{IC t interval} still hold. The insurer's profit from offering a contract $\left(R_{\theta},p_{\theta}\right)$ becomes 
\begin{equation}
\label{tilde pi}
\tilde\pi\left(R_{\theta},p_{\theta}\right)=\pi\left(R_{\theta},p_{\theta}\right)-k\cdot\One_{\left\{R_{\theta}(l)\not\equiv l \right\}},
\end{equation}

\noindent where $\pi$ is defined in \eqref{pi} and $p_\theta$ satisfies \eqref{IC t}.  Therefore,
the expected profit for the insurer is given by
\begin{align}
\label{VRT cost}
\tilde V\left(\left(R_{\theta}, p_{\theta}\right)_{\theta\in\Theta}\right)
=\int_\Theta
\tilde\pi\left(R_{\theta},p_{\theta}\right)\mathrm{d}F(\theta),
\end{align}

\noindent and the profit-maximizing, IR, and IC menus are obtained by solving the following problem.
\begin{problem}
\label{maxcost}
\[ (R^\ast_\theta,p^\ast_\theta)_{\theta\in\Theta}\in\underset{\left(R_\theta,p_\theta\right)_{\theta\in\Theta}\in\cIr\cap\cIc}\argmax \, \tilde V\left(\left(R_{\theta}, p_{\theta}\right)_{\theta\in\Theta}\right).\]
\end{problem} 

We solve this problem using a similar approach to that presented in Theorem \ref{profit solution}. First, we derive the expression for the insurer's profit with an IC menu, similar to Proposition \ref{IC menu}(2). The result is summarized below.

\begin{proposition}
\label{profit expression cost} 
If $\left(R_{\theta},p_{\theta}\right)_{\theta\in\Theta}\in\cIc,$ then the insurer's expected profit, as defined in \eqref{VRT cost}, is given by
\begin{align}
\label{profit cost}
\tilde V\left(\left(R_{\theta}, p_{\theta}\right)_{\theta\in\Theta}\right)
=V\left(\left(R_{\theta}, p_{\theta}\right)_{\theta\in\Theta}\right)
-k\cdot\mathbb{P}\left(\theta\notin\Theta_R\right),
\end{align}
where  $\Theta_R:=\left\{\theta\in\Theta\mid R_{\theta}(l)=l\,\text{for all}\,\, l\in[0,\overline{L}]\right\},$ and $V\left(\left(R_{\theta}, p_{\theta}\right)_{\theta\in\Theta}\right)$ is defined in \eqref{IC profit}.
\end{proposition}

Subsequently, replacing the expected cost in Problem \ref{maxcost} by the expression obtained in \eqref{profit cost}, we find the optimal insurance structure by pointwise maximization. The result is stated in the following theorem.

\begin{theorem}
\label{profit solution cost}
Suppose that $J_\theta(l)$ given in \eqref{virtual value J}
is non-decreasing in $\theta$, for all $l$. Then,
\begin{enumerate}
\item If $k>\displaystyle\int_0^{\overline{L}}\left[H(l)-g_{\overline\theta}\left(H\left(l\right)\right)\right]\mathrm{d}l,$  then  coverage is denied.

\item Otherwise, Problem \ref{maxcost} has a solution $\left(R^\ast_\theta,p^\ast_\theta\right)_{\theta\in\Theta}\in\cIr\cap\cIc$, where $R_\theta^\ast(l)\equiv l$ when $\theta<\theta^\ast,$ and $(R_\theta^\ast,p_\theta^\ast)$  is the same as the solution given in Theorem \ref{profit solution} when $\theta\geq\theta^\ast$, where,
$$\theta^\ast:=\max\left\{\theta\in\Theta
\ \Big\vert \ 
\int_0^{\overline{L}}J_\theta(l)\One_{\left\{J_\theta(l)>0\right\}}\mathrm{d}l\leq k\right\}.$$ 
Moreover, this menu is more profitable than a pooling equilibrium contract.
\end{enumerate}
\end{theorem}

Note that the cost threshold $\displaystyle\int_0^{\overline{L}}\left[H(l)-g_{\overline\theta}\left(H\left(l\right)\right)\right]\mathrm{d}l$ is the risk premium associated with the wealth of the type $\overline \theta$ consumer (the most risk-averse type) in the absence of insurance. Indeed,
\begin{align*}
\int_0^{\overline{L}}\left[H(l)-g_{\overline\theta}\left(H\left(l\right)\right)\right]\mathrm{d}l
&= -\int_0^{\overline{L}} \left[1-H(l)\right] \mathrm{d}l
+ \int_0^{\overline{L}} \left[1-g_{\overline\theta}(H(l))\right] \mathrm{d}l \\
&= \E[-L] - U_{\overline\theta}(L,0) 
=\E[-L] - DU_{\overline\theta}(-L)\\
&=\Delta_{\overline\theta}(-L) \ (> 0).
\end{align*}
This risk premium is non-decreasing with $\theta$, by Assumption \ref{gtheta}: for $\theta_1, \theta_2 \in \Theta$ such that 
$\theta_1 \leq \theta_2$, $g_{\theta_1}(s) \geq g_{\theta_2}(x)$, for all $x \in [0,1]$, i.e., $\theta_2$ is more risk averse than $\theta_1$. Consequently,
\begin{equation}
\label{TypeOrderProfit}
\int_0^{\overline{L}}\left[H(l)-g_{\theta}\left(H\left(l\right)\right)\right]\mathrm{d}l
\leq \int_0^{\overline{L}}\left[H(l)-g_{\overline\theta}\left(H\left(l\right)\right)\right]\mathrm{d}l, \ \ \forall \, \theta \in \Theta.
\end{equation}

In addition, $\displaystyle\int_0^{\overline{L}}\left(H(l)-g_{\overline\theta}\left(H\left(l\right)\right)\right)\mathrm{d}l$ is the expected profit of the insurer when there is no hidden information (with $\Theta=\{\overline{\theta}\}$) and no friction costs. Indeed, by Proposition \ref{one}, the optimal contract in that case is 
$$(R^*_{\overline\theta}, p^*_{\overline\theta}) 
= \left(0,\displaystyle\int_{0}^{\overline{L}}\left[1-g_{\overline\theta}\left(H\left(l\right)\right)\right]\mathrm{d}l\right),$$
which leads to an expected profit of 
\begin{align*}
\pi(R^*_{\overline\theta}, p^*_{\overline\theta})
&=\int_{0}^{\overline{L}}\left[1-g_{\overline\theta}\left(H\left(l\right)\right)\right]\mathrm{d}l
-\mathbb{E}[L]
+\int_0^{\overline{L}}\left[1-H\left(l\right)\right] \, (R^*_{\overline\theta})^\prime(l) \, \mathrm{d}l \\ 
&=\int_{0}^{\overline{L}}\left[1-g_{\overline\theta}\left(H\left(l\right)\right)\right]\mathrm{d}l
-\int_{0}^{\overline L}\left[1-H(l)\right] \, \mathrm{d}l\\
&=\int_0^{\overline{L}}\left[H(l)-g_{\overline\theta}\left(H\left(l\right)\right)\right]\mathrm{d}l.
\end{align*}

Theorem \ref{profit solution cost} states that if the fixed cost $k$ exceeds the profit that the insurer could have obtained in a market with perfect information and no friction cost, and in which the consumer is of the most risk-averse type $\overline\theta$, then the insurer anticipates losses even on the most risk-averse consumers (and hence on all other types $\theta$, by \eqref{TypeOrderProfit}). In that case, there will be no gains from trade, and the insurer is not willing to offer any contract. A similar conclusion is reached in \citet[Theorem 1]{CHADE2020105085}, which states that if there are no gains from trade at a given belief (the distribution of loss probabilities), then there are no gains from trade at any belief that likelihood-ratio dominates it (i.e., is more risky in the sense of the likelihood ratio).

On the other hand, when the fixed cost $k$ is less than this threshold $\int_0^{\overline{L}}\left[H(l)-g_{\overline\theta}\left(H\left(l\right)\right)\right]\mathrm{d}l$, Theorem \ref{profit solution cost} shows that the introduction of the fixed cost does not alter the equilibrium derived in Theorem \ref{profit solution}, where no friction costs were present, with the exception that certain contracts are excluded. Indeed, for risk types $\theta < \theta^*$, the equilibrium indemnification provides for no insurance coverage (full loss retention). This excluded part of the no-cost equilibrium of Theorem \ref{profit solution} can be interpreted as a pooling at no-coverage. Note that this exclusion occurs from below,  meaning that the contracts removed are those designed for the least risk-averse consumers, who are in fact less valuable for the insurer. A similar effect is discussed in \citet[Theorem 3 and Corollary 1]{CHADE2020105085}, which demonstrates that when a fixed cost is introduced, only pooling contracts and no-coverage contracts remain. However, contrary to \cite{CHADE2020105085}, Theorem \ref{profit solution cost} suggest that when trade occurs, separating equilibrium contracts always outperform pooling equilibrium contracts, despite the inclusion of fixed costs.

\subsection{Equilibria with Costs and Full Information}\vspace{-0.25cm}

If the insurer is able to observe the risk type of the consumer, say $\Theta=\{\theta_0\}$, then the introduction of the fixed cost $k>0$ has important implications on the full-information optimal contract derived in Proposition \ref{one}.

\begin{corollary}
\label{one cost}
If $\theta\equiv\theta_0,$ then an optimal contract is given by the following: 
\begin{enumerate} 
\item If $k>\int_{0}^{\overline{L}}\left[H\left(l\right)-g_{\theta_0}\left(H\left(l\right)\right)\right]\mathrm{d}l$, then coverage is denied. That is, 
$$(R^*_{\theta_0}, p^*_{\theta_0}) = \left(L,0\right).$$

\item If $k\leq \int_{0}^{\overline{L}}\left[H\left(l\right)-g_{\theta_0}\left(H\left(l\right)\right)\right]\mathrm{d}l,$ then full insurance is optimal, at a premium that makes the consumer indifferent between insurance and no-insurance. That is,  
$$(R^*_{\theta_0}, p^*_{\theta_0}) = \left(0,\displaystyle\int_{0}^{\overline{L}}\left[1-g_{\theta_0}\left(H\left(l\right)\right)\right]\mathrm{d}l\right).$$
\end{enumerate}
\end{corollary}

Corollary \ref{one cost} suggests that if the fixed cost $k$ exceeds the profit that the insurer could have obtained in the absence of friction costs, then the insurer anticipates losses and will not provide any coverage. However, when the fixed cost is lower than the threshold $\int_{0}^{\overline{L}}\left[H\left(l\right)-g_{\theta_0}\left(H\left(l\right)\right)\right]\mathrm{d}l$, the optimal coverage is similar to the no-cost full-information optimum of  Proposition \ref{one}.

\section{Pareto-Optimal Menus of Contracts}
\vspace{-0.25cm}
\label{SecPO}

The concept of incentive-compatible Pareto-optimality (often referred to as incentive efficiency) in the context of adverse selection has been examined by \cite{Prescotti1984}, \cite{CrockerSnow85}, 
\cite{DASPREMONT1990233}, \cite{Jerez2003}, and \cite{BisinGottardi2006}, for instance, but with finitely many types. To define incentive-efficient menus of contracts in a context of an arbitrary (unaccountably infinite) set of types $\Theta$, we use the approach to perfect competition in large economies introduced by \cite{Aumann1964}, and in particular the notion of a continuum of consumers, or small traders in the language of \cite{Aumann1964} and \cite{Shitovitz1973}.

\subsection{Incentive-Efficient Menus}\vspace{-0.25cm}

Equip $\Theta = \left[\underline \theta, \overline \theta\right]$ with the Borel sigma-algebra $\Sigma_\mathcal{B}(\Theta)$ and Lebesgue measure $\mathcal{L}$. A probability measure $\eta_1$ on $\left(\Theta, \Sigma_\mathcal{B}(\Theta)\right)$ is said to be absolutely continuous with respect to a probability measure $\eta_2$ on $\left(\Theta, \Sigma_\mathcal{B}(\Theta)\right)$, if $\eta_2(B) = 0 \Longrightarrow \eta_1(B)=0$, for each $B \in \Sigma_\mathcal{B}(\Theta)$. If $\eta_1$ is absolutely continuous with respect to $\eta_2$ and $\eta_2$ is absolutely continuous with respect to $\eta_1$, the two measures are said to be equivalent, and we write $\eta_1 \sim \eta_2$.

Let $\mu$ be a probability measure on $\left(\Theta, \Sigma_\mathcal{B}(\Theta)\right)$ that represents the distribution of consumer types in the market. Since $\mu(\Theta) = 1$ by normalization, one can think of the quantity $\mu(B)$ as representing the proportion of types in the set $B \in \Sigma_\mathcal{B}(\Theta)$ in the market. For each $\theta \in \Theta$, let $F(\theta) := \mu\left([\underline \theta, \theta]\right)$ define the cumulative distribution function over types, and let $\overline F(\theta) := \mu\left((\theta,\overline \theta]\right)$ define the decumulative distribution function over types. We assume that $\mu$ is absolutely continuous with respect to $\mathcal{L}$, with Radon–Nikodym derivative given by the probability density function $f$ of the distribution function $F$ over types.

\begin{definition}
A menu $\left(R^\ast_{\theta},p^\ast_{\theta}\right)_{\theta\in\Theta}\in\cIr\cap\cIc$ is said to be incentive efficient, or incentive-Pareto-optimal (IPO), if there does not exist any menu $\left(R_{\theta},p_{\theta}\right)_{\theta\in\Theta}\in\cIr\cap\cIc$ such that the following two conditions hold:
\begin{enumerate}
\item $U_\theta\left(R_{\theta},p_{\theta}\right)
\geq
U_\theta\left(R^\ast_{\theta},p^\ast_{\theta}\right)$, for all $\theta \in \Theta$, 
and 
$\displaystyle\int_\Theta
\pi\left(R_{\theta},p_{\theta}\right)\mathrm{d}F(\theta) \geq \int_\Theta
\pi\left(R^\ast_{\theta},p^\ast_{\theta}\right)\mathrm{d}F(\theta)$, with at least one strict inequality.

\item If $\displaystyle\int_\Theta
\pi\left(R_{\theta},p_{\theta}\right)\mathrm{d}F(\theta) = \int_\Theta
\pi\left(R^\ast_{\theta},p^\ast_{\theta}\right)\mathrm{d}F(\theta)$, then the set of types for which the inequalities are strict is non-null. That is,
$$\mu \bigg(\Big\{\theta \in \Theta: U_{\theta}\left(R_{\theta},p_{\theta}\right)
>
U_{\theta}\left(R^\ast_{\theta},p^\ast_{\theta}\right)\Big\}\bigg) > 0.$$
\end{enumerate}

We denote by $\mathcal{IPO} \subseteq \mathcal{IR \cap \mathcal{IC}}$ the set of all IPO menus.
\end{definition}

\noindent Equivalently, one can drop condition $(2)$ from the above definition, with the understanding that statements made for all $\theta$ are to be taken in the $\mu$-a.s.\ sense, as is common in the literature on markets with a continuum of traders (e.g., the convention used in \cite{Aumann1966} and the subsequent literature).

\begin{proposition}
\label{ProbWFmaxPO}
If there exists a probability measure $\eta$ on $\left(\Theta, \Sigma_\mathcal{B}(\Theta)\right)$ such that $\mu$ is absolutely continuous with respect to $\eta$, 
and some $\alpha \in (0,1)$, such that $(R^\ast_{\eta,\alpha,\theta},p^\ast_{\eta,\alpha,\theta})_{\theta\in\Theta}$ is optimal for the problem
\begin{equation}
\label{POmax2}
\underset{\left(R_{\theta},p_{\theta}\right)_{\theta\in\Theta}\in\cIr\cap\cIc} \sup \,
\left[ \alpha \,
\int_\Theta U_\theta\left(R_{\theta},p_{\theta}\right)  \, \mathrm{d}\eta
+
(1-\alpha) \,
\int_\Theta
\pi\left(R_{\theta},p_{\theta}\right)\,\mathrm{d}\mu
\right],
\end{equation}
then $(R^\ast_{\eta,\alpha,\theta},p^\ast_{\eta,\alpha,\theta})_{\theta\in\Theta} \in \mathcal{IPO}$.
\end{proposition}

\subsection{Characterizing Incentive-Efficient Menus}\vspace{-0.25cm}

By Proposition \ref{ProbWFmaxPO}, we focus on maximizing the social welfare function
\begin{equation}
\label{EqSocialWelfareFct}
\begin{split}
W_{\eta,\alpha}\left(\left(R_{\theta}, p_{\theta}\right)_{\theta\in\Theta}\right) 
&:= 
\alpha \,
\int_\Theta U_\theta\left(R_{\theta},p_{\theta}\right)  \, \mathrm{d}\eta
+
(1-\alpha) \,
\int_\Theta
\pi\left(R_{\theta},p_{\theta}\right)\,\mathrm{d}\mu,
\end{split}
\end{equation}
for a given probability measure $\eta$ on $\left(\Theta, \Sigma_\mathcal{B}(\Theta)\right)$, and  a given weight $\alpha \in (0,1)$: 
\begin{problem}
\label{max po}
\begin{equation}
\underset{\left(R_\theta,p_\theta\right)_{\theta\in\Theta}\in\cIr\cap\cIc} \sup \,
W_{\eta,\alpha}\left(\left(R_{\theta}, p_{\theta}\right)_{\theta\in\Theta}\right).
\end{equation}
\end{problem}

Solutions to Problem \ref{max po} are difficult to characterize in general. Hereafter, we make some assumptions on $\eta$ that allow us to provide a crisp characterization of optima. First, let $F_\eta$ and $\overline{F}_\eta$ denote, respectively, the cumulative distribution function and survival function over types induced by $\eta$, that is, $F_\eta(\theta):=\eta([\underline\theta,\theta])$ and $\overline{F}_\eta(\theta):=\eta((\theta,\overline\theta])$, for all $\theta\in\Theta$. We will assume that $\eta$ and $\mu$ are equivalent, which ensures the existence of a probability density function $f_\eta$ for $F_\eta$, and that $\mu$ is smaller than  $\eta$ in the hazard rate order:   

\begin{assumption}
\label{InEta} 
The measure $\eta$ satisfies the following:
\begin{enumerate}
\item $\eta \sim \mu$.
\item $\frac{f(\theta)}{\overline{F}(\theta)}\geq\frac{f_\eta(\theta)}{\overline{F}_\eta(\theta)}$, for all $\theta \in \Theta$.
\end{enumerate}
\end{assumption}

\begin{remark}\label{EtaAndMu}
Assumption \ref{InEta}(2) implies that $\frac{\overline{F}_\eta(\theta)}{\overline{F}(\theta)}$ is non-decreasing in $\theta \in \Theta.$ Indeed, 
\begin{align*}
\left(\frac{\overline{F}_\eta(\theta)}{\overline{F}(\theta)}\right)'=\frac{-f_\eta(\theta)\overline{F}(\theta)+\overline{F}_\eta(\theta)f(\theta)}{\overline{F}(\theta)^2}
=\frac{\overline{F}_\eta(\theta)}{\overline{F}(\theta)}
\left(\frac{f(\theta)}{\overline{F}(\theta)}
-
\frac{f_\eta(\theta)}{\overline{F}_\eta(\theta)}
\right)\geq0, \ \ \forall \theta \in \Theta.
\end{align*}

\noindent Moreover, since $\frac{\overline{F}_\eta(\theta)}{\overline{F}(\theta)}\bigg\vert_{\theta=\underline\theta}=1,$ and 
\[\frac{\overline{F}_\eta(\theta)}{\overline{F}(\theta)}\bigg\vert_{\theta=\overline\theta}=\lim\limits_{\theta\rightarrow\overline\theta}\frac{\overline{F}_\eta(\theta)}{\overline{F}(\theta)}=\lim\limits_{\theta\rightarrow\overline\theta}\frac{-f_\eta(\theta)}{-f(\theta)}
=\frac{f_\eta(\overline\theta)}{f(\overline\theta)},\]
where the second equality is by  L'Hospital's rule, it follows that 
$\frac{f_\eta(\overline\theta)}{f(\overline\theta)}\geq 1$, or $f_\eta(\overline\theta)\geq f(\overline\theta)$, implying that $\frac{f(\overline{\theta})}{f(\overline{\theta})+f_\eta(\overline{\theta})} \leq \frac{1}{2}$.
\end{remark}

Under Assumption \ref{InEta}, the following result provides a characterization of solutions of Problem \ref{max po}.

\begin{theorem}
\label{profit solution pareto}
Under Assumption \ref{InEta}, an optimal solution to Problem \ref{max po} is characterized as follows:
\begin{enumerate}
\item If $\alpha\in\left(0,\frac{f(\overline{\theta})}{f(\overline{\theta})+f_\eta(\overline{\theta})}\right)$  and if the function $J_{\eta,\alpha,\theta}(l)$ given in \eqref{J eta alpha theta} is non-decreasing in $\theta$ for all $l$, then there exists a solution  $\left(R^\ast_{\eta,\alpha,\theta},p^\ast_{\eta,\alpha,\theta}\right)_{\theta\in\Theta}$. For each $\theta \in \Theta$, the optimal retention function is such that the marginal retention satisfies the following:
\begin{equation}
\label{band-band pareto}
\frac{\partial R^\ast_{\eta,\alpha,\theta}(l)}{\partial l}=
\left\{
\begin{array}[c]{ll}%
0, &  J_{\eta,\alpha,\theta}(l)>0,\vspace{0.2cm}\\
\in[0,1], &  J_{\eta,\alpha,\theta}(l)=0,\vspace{0.2cm}\\
1, & J_{\eta,\alpha,\theta}(l)< 0,
\end{array}
\right.  
\end{equation}
and the premia $\left\{p^*_{\eta,\alpha,\theta}\right\}_{\theta \in \Theta}$ satisfy \eqref{optimal t}.
The function $J_{\eta,\alpha,\theta}(l)$ is defined as
\begin{align}\label{J eta alpha theta}
    J_{\eta,\alpha,\theta}(l)
    &:=\left(1-\alpha\right)\left[H(l)-g_{\theta}\left(H\left(l\right)\right)\right]
    \,+\,\frac{\overline{F}(\theta)}{f(\theta)}
    \left(1-\alpha-\alpha\frac{\overline{F}_\eta(\theta)}{\overline{F}(\theta)}\right)\, \frac{\partial g_\theta\left(H\left(l\right)\right)}{\partial\theta}.
\end{align}

\item If $\alpha\in\left[\frac{f(\overline{\theta})}{f(\overline{\theta})+f_\eta(\overline{\theta})},\frac{1}{2}\right]$ and if the function $J_{\eta,\alpha,\theta}(l)$ given in \eqref{J eta alpha theta} is non-decreasing in $\theta$ for all $l$ when $\theta<\theta_\alpha,$ then $R^\ast_{\eta,\alpha,\theta}$ follows the form given in \eqref{band-band pareto} for $\theta<\theta_\alpha,$ and $R^\ast_{\eta,\alpha,\theta}=0$  for $ \theta\geq\theta_\alpha$, where $\theta_\alpha$ is determined by the equation   $\frac{\overline{F}_\eta(\theta_\alpha)}{\overline{F}(\theta_\alpha)}=\frac{1-\alpha}{\alpha}$. The premia $\left\{p^*_{\eta,\alpha,\theta}\right\}_{\theta \in \Theta}$ satisfy \eqref{optimal t}.

\item If $\alpha\in\left(\frac{1}{2},1\right),$  then $R^\ast_{\eta,\alpha,\theta}=0$ and $p^*_{\eta,\alpha,\theta}=0$, for all $ \theta\in\Theta$.
\end{enumerate}

\noindent Moreover, the collection $\left\{R^*_{\eta,\alpha,\theta}\right\}_{\theta \in \Theta}$ of optimal retention functions is submodular, for a given $\eta$ and $\alpha$.
\end{theorem}

This result characterizes incentive-efficient contracts as the solutions to a social welfare maximization problem, where the consumer's welfare is weighted by $\alpha$, and $\eta$ can be seen as the social planner's prior over unknown types. The structure of incentive-efficient contracts  depends on the value of $\alpha$. When $\alpha\in\left(0,\frac{f(\overline{\theta})}{f(\overline{\theta})+f_\eta(\overline{\theta})}\right)$,  a separating equilibrium emerges,  and the optimal retention function follows the submodular layered structure described in \eqref{band-band pareto}, ensuring that higher-risk types receive more coverage, similarly to Theorem \ref{profit solution}.  When $\alpha\in\left[\frac{f(\overline{\theta})}{f(\overline{\theta})+f_\eta(\overline{\theta})},\frac{1}{2}\right],$ the equilibrium remains separating, but with a pooling segment at the top: full insurance is offered to all consumers with type $\theta\geq\theta_\alpha$. Moreover, $\theta_\alpha$ decreases with $\alpha$ since $\theta_\alpha$ is the solution to $\frac{\overline{F}_\eta(\theta)}{\overline{F}(\theta)}=\frac{1-\alpha}{\alpha}$, where the fraction $\frac{1-\alpha}{\alpha}$ decreases with $\alpha,$ while the ratio $\frac{\overline{F}_\eta(\theta)}{\overline{F}(\theta)}$ increases with $\theta$. This implies that as $\alpha$ increases, more consumer types become pooled at the top.
When  $\alpha>\frac{1}{2}$,  a pooling equilibrium arises. That is, if a sufficiently high weight is placed on consumer welfare, the previously separating equilibrium transitions into a pooling equilibrium, where full insurance is provided to all consumers. This result is at odds with the findings in the base cases discussed in Sections \ref{Profit-O-menu} and \ref{SecCost}, where a separating structure is always preferred over a pooling structure in terms of optimal contract design.

To achieve the separating layered equilibrium when $\alpha\leq\frac{1}{2}$, we impose a monotonicity condition on the  function $J_{\eta,\alpha,\theta}$,  analogous to the condition on $J_\theta$ in Theorem \ref{profit solution}. These two functions are closely related, as expressed in the following relationship:
   \begin{align}\label{J and J alpha} J_{\eta,\alpha,\theta}(l)
   =(1-\alpha) \, J_\theta(l)
   \,-\,
   \alpha\,\left(\frac{\overline{F}(\theta)}{f(\theta)}\right) \left(\frac{\overline{F}_\eta(\theta)}{\overline{F}(\theta)}\right)\,\frac{\partial g_\theta\left(H\left(l\right)\right)}{\partial \theta}. \end{align}
   In particular, when $\alpha = 0$, $J_{\eta,\alpha,\theta}(l)$ reduces to $J_{\theta}(l)$, recovering the standard case without explicit consumer welfare considerations.
The optimal retention function structure in \eqref{band-band pareto} resembles the solution in \eqref{band-band}, with the key distinction that the layer formation now depends on 
 $J_{\eta,\alpha,\theta}(l)$ rather than $J_{\theta}(l)$.
   When this function is non-decreasing in $\theta$ for a given $l$, the existence of an optimal retention function is ensured. Furthermore, the condition guaranteeing the monotonicity of $J_{\theta}(l)$ in $\theta$, as discussed in Section \ref{monotonicity},  is also sufficient to ensure the monotonicity of  $J_{\eta,\alpha,\theta}(l)$ in $\theta$. Specifically,
\begin{align*}
\frac{\partial J_{\eta,\alpha,\theta}(l)}{\partial \theta}
&=-\frac{\partial g_\theta\left(H\left(l\right) \right)}{\partial \theta}\left((1-\alpha)\theta-\frac{\overline{F}(\theta)}{f(\theta)}\left(1-\alpha-\alpha\frac{\overline{F}_\eta(\theta)}{\overline{F}(\theta)}\right)\right)^\prime\\
&\quad\quad\quad\quad\quad\quad+\frac{\overline{F}(\theta)}{f(\theta)}\left(1-\alpha-\alpha\frac{\overline{F}_\eta(\theta)}{\overline{F}(\theta)}\right)\frac{\partial^2 g_\theta\left(H\left(l\right)\right)}{\partial \theta^2}.
\end{align*}

\noindent A non-decreasing hazard rate   $\nicefrac{f}{\overline{F}}$ under $\mu$,  the convexity of $\theta \mapsto g_\theta(s)$  for each  $s\in(0,1)$, and Assumption \ref{InEta} together ensure that $J_{\eta,\alpha,\theta}(l)$ is non-decreasing in $\theta$ for  all $l$  when $\alpha\in\left(0,\frac{f(\overline{\theta})}{f(\overline{\theta})+f_\eta(\overline{\theta})}\right),$ and $J_{\eta,\alpha,\theta}(l)$ is non-decreasing in $\theta$ when $\theta<\theta_\alpha$ for all $l$ when $\alpha\in\left[\frac{f(\overline{\theta})}{f(\overline{\theta})+f_\eta(\overline{\theta})},\frac{1}{2}\right].$ These results follow from the fact that  $\frac{\overline{F}_\eta(\theta)}{\overline{F}(\theta)}$ increases with $\theta$ and, in these cases, $1-\alpha-\alpha\frac{\overline{F}_\eta(\theta)}{\overline{F}(\theta)}\geq0.$

\medskip

The following proposition outlines key properties of the IPO menu established in Theorem \ref{profit solution pareto}.

\begin{proposition}
\label{profit solution pareto properties}
Under the same assumption as Theorem \ref{profit solution pareto}, the following hold:
\begin{enumerate}
\item For each $\theta\in\Theta,$ the marginal function $\frac{\partial R_{\eta,\alpha,\theta}^{\ast}(l)}{\partial l}$ decreases with $\alpha$ for all $l$. Consequently,  the optimal retention function  $R_{\eta,\alpha,\theta}^{\ast}(l)$ decreases with $\alpha.$
\item For each $\theta\in\Theta$, $U_\theta(R_{\eta,\alpha,\theta}^{\ast},p^\ast_{\eta,\alpha,\theta})$ increases with $\alpha.$
\item The insurer's total profit $V\big((R^\ast_{\eta,\alpha,\theta}, p^\ast_{\eta,\alpha,\theta})_{\theta\in\Theta}\big)$ decreases with $\alpha.$ Moreover, $V\big((R^\ast_{\eta,\alpha,\theta}, p^\ast_{\eta,\alpha,\theta})_{\theta\in\Theta}\big) > 0$, for  $\alpha\in\left(0,\frac{1}{2}\right]$. 
\end{enumerate}
\end{proposition}

The first result shows that the IPO menu provides greater insurance coverage for each consumer type, with the retained losses always decreasing as more weight is placed on consumer welfare. Second, the incentive efficient menu, which differs in coverage and pricing structure from the profit-maximizing contract, benefits all consumer types, with utility levels increasing monotonically with $\alpha$. In contrast, the insurer's total profit decreases  with $\alpha$. The derived contract menu may become unprofitable for the insurer when consumer welfare is heavily prioritized, particularly when $\alpha>\frac{1}{2}$. However, for $\alpha\leq\frac{1}{2}$, the insurer's profit remains strictly positive under this IPO menu.

\subsection{Pareto Optima with Full Information}\vspace{-0.25cm}

If the insurer is able to observe the risk type of the consumer, say $\Theta=\{\theta_0\}$, then there is no adverse selection or screening problem. In this case, it is easy to verify that a contract $(R,p) \in \mathcal{R}_L \times \bbR_+$ is Pareto optimal if and only if there is some $\alpha \in [0,1]$ such that $(R,p)$ is optimal for the following problem:
\begin{problem}
\label{max3}
$$\underset{(R,p) \in \mathcal{R}_L \times \bbR_+} \sup \, \Big\{\alpha\, U_{\theta_0}(R,p)+(1-\alpha)\,\pi(R,p): U_{\theta_0}(R,p) \geq U_{\theta_0}(L,0)\Big\}.$$
\end{problem}

The following result shows that under complete information, full insurance is provided to the consumer, as in Proposition \ref{one}, while the premium depends on the weight $\alpha$.

\begin{corollary}\label{one Pareto}
If $\Theta = \left\{\theta_0\right\}$, an optimal solution to Problem \ref{max3} is given by the following:
\begin{enumerate}
\item If $\alpha\leq\frac{1}{2},$ then $$(R^*_{\theta_0}, p^*_{\theta_0}) = \left(0,\displaystyle\int_0^{\overline{L}}\left[1-g_{\theta_0}\left(H\left(l\right)\right)\right]\mathrm{d}l\right).$$

\item If $\alpha>\frac{1}{2},$ then $$(R^*_{\theta_0}, p^*_{\theta_0}) = \left(0,0\right).$$
\end{enumerate}
\end{corollary}

This result states that when $\alpha$ is small, the insurer sets the premium in a way to extract all of the consumer surplus, yielding a strictly positive profit of $\Delta_{\theta_0}(-L)$. However, when consumer welfare is prioritized (i.e., $\alpha$ is large), the premium is reduced, and  the insurer charges a zero premium.


\section{Conclusion}\vspace{-0.25cm}
\label{SecConc}

This paper examines a monopolistic insurance market in which consumers have private information about their risk attitudes, while the insurer observes only the overall risk distribution. This hidden information is modeled using the Dual Utility model of \cite{yaari}, and equilibrium contracts are designed to maximize the insurer’s profit while ensuring incentive compatibility—that is, revealing consumers' private information. Our model extends the classical setting of \citet{stiglitz1977} by incorporating a continuum of types and assuming that losses are continuously distributed. Under certain conditions, a separating equilibrium is optimal and can be fully characterized. The continuum framework allows for a detailed analysis of key properties of the optimal contract, including consumer welfare and insurer profit.

We characterize the optimal (profit-maximizing) incentive-compatible and individually rational menus of insurance contracts in terms of the marginal loss retention per type of consumer. Optima consist of layered deductible contracts, where each such layered structure is determined by the risk type of the consumer, proxied by their probability weighting function. Such layered indemnity structures are widely observed in practice. The key properties of our separating equilibria are that (i) insurance coverage and premia are monotone in the level of risk aversion; (ii) the most risk-averse consumer receives full insurance (\textit{efficiency at the top}); (iii) the monopoly will absorb all surplus from the least-risk averse consumer, and the contract offered to this consumer leaves them indifferent between participation and non-participation; (iv) at an optimum, the consumer's utility of wealth is a decreasing function of their risk type; and (v) consumers with a higher level of risk aversion, who are willing to pay more for a product with a higher coverage, are generally more valuable to the insurer, in that they induce a higher expected profit for the insurer.

In addition, we examine the effect of fixed insurance provision costs on equilibria. We show that the optimal menu is the same as in the absence of such  fixed costs, with the exception that only part of the menu is excluded. The excluded contracts are those designed for consumers with relatively lower risk aversion, who are less valuable to the insurer.  Moreover, if the fixed cost is prohibitively high, exceeding the profit that the insurer could have obtained in a market with perfect information and no friction cost, and in which the consumer is of the most risk-averse type, then the insurer anticipates losses even on the most risk-averse consumers (and hence on all other types). In that case, there will be no gains from trade, and the insurer is not willing to offer any contract. However, when trade occurs, separating equilibrium contracts always outperform pooling equilibrium contracts.

Finally, we characterize incentive-efficient menus of contracts in the context of an arbitrary type space. We show that individually rational and incentive compatible contracts that are Pareto optimal can be achieved by maximizing a social welfare function that accounts for hidden types. While it is difficult to solve such a problem in the general case, we are able to provide a crisp characterization of solutions under a few assumptions.

An interesting direction for future research is to extend the current continuum-type model to a multidimensional hidden information framework in a monopoly market—specifically, by incorporating both hidden risk distributions and hidden risk attitudes. The analysis of discrete-type consumers, particularly the two-type case, has been explored in the literature. For instance, \citet{LANDSBERGER1994392} considers only hidden risk attitudes, while \citet{landsberger1999general} accounts for both hidden risk distributions and risk attitudes. Their findings suggest that when consumers differ only in risk attitudes, full insurance is offered to the more risk-averse group, whereas the less risk-averse type receives a two-valued partial insurance contract. When consumers also differ in risk distributions, the consumer with the lower certainty equivalent is offered full insurance, while the other type receives partial insurance, with the level of coverage depending on the ratio of the two risk distribution functions. Whether these conclusions extend to the continuum-type setting remains an open question. In the one-dimensional screening problem, transitioning from discrete to continuous types reveals interesting structural properties of the contract menu, such as the convexity of consumer welfare and the monotonicity of insurer profitability. However, when consumers also differ in risk distributions, it remains unclear whether these key properties hold. Additionally, it is important to examine whether the separating equilibrium continues to be more profitable and whether the equilibrium contract depends on the relative ordering of consumers’ risk distributions, as discussed in \citet{landsberger1999general}.

Another promising avenue for future research is the study of equilibrium outcomes in different types of insurance markets. According to \citet{RothschildandStiglitz1976} and \citet{stiglitz1977}, differences in risk attitudes do not affect equilibrium contracts in a competitive market, whereas they do in a monopoly market. However, there has been little discussion of whether such differences influence equilibrium contracts in market structures that fall between these two extremes. In such cases, insurers may have heterogeneous beliefs about the distribution of consumer types. For example, in a duopoly, insurer $1$ might assign a distribution $F_1$ over types, whereas insurer $2$ might assign a distribution $F_2$ over types, even though the true distribution might be $F$. These discrepancies in beliefs could influence contract design and lead to strategic interactions between insurers, potentially resulting in a market equilibrium that differs from both the competitive and monopoly cases.


\clearpage
\newpage
\hypertarget{LinkToAppendix}{\ }
\appendix

\vspace{-0.8cm}

\renewcommand{\thesection}{Appendix }
\renewcommand{\theequation}{\Alph{section}.\arabic{equation}}
\renewcommand{\thetheorem}{\Alph{section}.\arabic{theorem}}
\renewcommand{\thecorollary}{\Alph{section}.\arabic{corollary}}

\section{}
\vspace{-0.25cm}
\subsection*{Proof of Proposition \ref{IRmenu}:}\vspace{-0.25cm}

\begin{proof*}
Consider a contract $\left(R_{\theta},p_{\theta}\right)_{\theta\in\Theta} \in \mathcal{R}_L^\Theta \times \mathbb{R}_+^\Theta$. If $\left(R_{\theta},p_{\theta}\right)_{\theta\in\Theta} \in \mathcal{IR}$, then for each $\theta \in \Theta$, 
\begin{align*}
-\int_0^{\overline{L}}\left[1-g_\theta\left(H\left(l\right)\right)\right]\mathrm{d}l
= U_\theta\left(L,0\right)
&\leq U_\theta\left(R_\theta,p_\theta\right)\\
&=-p_\theta-\int_0^{\overline{L}}\left[1-g_\theta\left(H\left(l\right)\right)\right]\frac{\partial R_{\theta}(l)}{\partial l}\mathrm{d}l,
\end{align*}
implying that
$$p_\theta 
\leq 
\int_0^{\overline{L}}\left[1-g_{\theta}(H(l))\right]\left(1-\frac{\partial R_{\theta}(l)}{\partial l}\right)\mathrm{d}l.$$
The converse follows similarly. 
\end{proof*}

\subsection*{Proof of Proposition \ref{IC menu}:}\vspace{-0.25cm}

\begin{proof*}
The utility for consumer of type $\theta$ with contract $\left(R_{\theta},p_{\theta}\right)$ in \eqref{Ugamma} can be written as
\begin{align}
\label{U expression}
U_\theta\left(R_{\theta},p_{\theta}\right)&=-p_\theta-\int_{0}^{R_\theta(\overline{L})}1-g_\theta\left(\mathbb{P}\left[R_{\theta}(L)\leq l\right]\right)\mathrm{d}l\nonumber\\
&=-p_\theta-\int_0^{R_\theta(\overline{L})}1-g_\theta\left(H\left(R^{-1}_{\theta}(l)\right)\right)\mathrm{d}l\nonumber\\
&=-p_\theta-\int_0^{\overline{L}}1-g_\theta\left(H\left(l\right)\right)\mathrm{d}R_{\theta}(l)\nonumber\\
&=-p_\theta-\int_0^{\overline{L}}\left[1-g_\theta\left(H\left(l\right)\right)\right]\frac{\partial R_{\theta}(l)}{\partial l}\mathrm{d}l,
\end{align} 
where $R_\theta^{-1}(l)=\sup\{z:R_\theta(z)\leq l\}.$ Similarly, for the contract $\left(R_{\theta'},p_{\theta'}\right),$ we have,
\[U_{\theta}\left(R_{\theta'},p_{\theta'}\right)=p_{\theta'}-\int_0^{\overline{L}}\left[1-g_\theta\left(H\left(l\right)\right)\right]\frac{\partial R_{\theta'}(l)}{\partial l}\mathrm{d}l.\]
Note that $U_{\theta}\left(R_{\theta'},p_{\theta'}\right)$ is Lipschitz continuous in $\theta$, and hence absolutely continuous in $\theta$, since 
\[\Bigg|\frac{\partial U_{\theta}\left(R_{\theta'},p_{\theta'}\right)}{\partial \theta}\Bigg|
\leq
\int_0^{\overline{L}}\Bigg|\frac{\partial g_\theta\left(H\left(l\right)\right)}{\partial \theta}\frac{\partial R_{\theta'}(l)}{\partial l}\Bigg|\mathrm{d}l
\leq
\int_0^{\overline{L}}\Bigg|\frac{\partial g_\theta\left(H\left(l\right)\right)}{\partial \theta}\Bigg|\mathrm{d}l
\leq c \, \overline L
<+\infty,\]

\noindent where the second inequality holds since $R_\theta\in\mathcal{R}_L$, and the third inequality holds by Assumption \ref{gtheta}(1).

Then by the envelope theorem (e.g., \citet[Theorem 2]{milgrom2002envelope}), if $\left(R_{\theta},p_{\theta}\right)_{\theta\in\Theta}\in\cIc$, then for any $\theta\in \Theta,$
\begin{align}
\label{IC utility}
U_\theta\left(R_{\theta},p_{\theta}\right)
&=U_{\underline\theta}\left(R_{\underline\theta},p_{\underline\theta}\right)+\int^\theta_{\underline{\theta}}\left.\frac{\partial U_{s'}(R_{s},p_s)}{\partial s'}\right\vert_{s'=s}\mathrm{d}s\nonumber\\
&=-p_{\underline\theta}-\int_0^{\overline{L}}\left[1-g_{\underline\theta}\left(H\left(l\right)\right)\right]\frac{\partial R_{\underline\theta}(l)}{\partial l}\mathrm{d}l+\int^\theta_{\underline{\theta}}\int_0^{\overline{L}}\frac{\partial g_s\left(H\left(l\right)\right)}{\partial s}\frac{\partial R_{s}(l)}{\partial l}\mathrm{d}l\,\mathrm{d}s.
\end{align}

\noindent Equating \eqref{U expression} and \eqref{IC utility}, it follows that 
\begin{align*}
p_\theta
&=p_{\underline\theta}+\int_0^{\overline{L}}\left[1-g_{\underline\theta}\left(H\left(l\right)\right)\right]\frac{\partial R_{\underline\theta}(l)}{\partial l}\mathrm{d}l-\int^\theta_{\underline{\theta}}\int_0^{\overline{L}}\frac{\partial g_s\left(H\left(l\right)\right)}{\partial s}\frac{\partial R_{s}(l)}{\partial l}\mathrm{d}l\mathrm{d}s-\int_0^{\overline{L}}\left[1-g_\theta\left(H\left(l\right)\right)\right]\frac{\partial R_{\theta}(l)}{\partial l}\mathrm{d}l.
\end{align*} 

\noindent Substituting the above expression for the premium $p_\theta$ into \eqref{pi}, the insurer's expected profit given in \eqref{VRT} becomes
\begin{align*}
V\left(\left(R_{\theta}, p_{\theta}\right)_{\theta\in\Theta}\right)
&=\int^{\overline{\theta}}_{\underline{\theta}}
\pi\left(R_{\theta},p_{\theta}\right)f(\theta)\mathrm{d}\theta\\
&=p_{\underline\theta}+\int_0^{\overline{L}}\left[1-g_{\underline\theta}\left(H\left(l\right)\right)\right]\frac{\partial R_{\underline\theta}(l)}{\partial l}\mathrm{d}l-\mathbb{E}[L]\\
&-\int_{\underline{\theta}}^{\overline{\theta}}\left(\int^\theta_{\underline{\theta}}\int_0^{\overline{L}}\frac{\partial g_s\left(H\left(l\right)\right)}{\partial s}\frac{\partial R_{s}(l)}{\partial l}\mathrm{d}l\mathrm{d}s+\int_0^{\overline{L}}\left[H(l)-g_\theta\left(H\left(l\right)\right)\right]\frac{\partial R_{\theta}(l)}{\partial l}\mathrm{d}l\right)f(\theta)\mathrm{d}\theta\\
&=p_{\underline\theta}+\int_0^{\overline{L}}\left[1-g_{\underline\theta}\left(H\left(l\right)\right)\right]]\frac{\partial R_{\underline\theta}(l)}{\partial l}\mathrm{d}l-\mathbb{E}[L]-\int_{\underline{\theta}}^{\overline{\theta}}\left(\int_0^{\overline{L}}\left[H(l)-g_\theta\left(H\left(l\right)\right)\right]\frac{\partial R_{\theta}(l)}{\partial l}\mathrm{d}l\right)f(\theta)\mathrm{d}\theta\\
&+\left.\left(\int^\theta_{\underline{\theta}}\int_0^{\overline{L}}\frac{\partial g_s\left(H\left(l\right)\right)}{\partial s}\frac{\partial R_{s}(l)}{\partial l}\mathrm{d}l\mathrm{d}s\overline{F}(\theta)\right)\right\vert_{\theta=\underline{\theta}}^{\theta=\overline{\theta}}-\int_{\underline{\theta}}^{\overline{\theta}}\overline{F}(\theta)\int_0^{\overline{L}}\frac{\partial g_\theta\left(H\left(l\right)\right)}{\partial \theta}\frac{\partial R_{\theta}(l)}{\partial l}\mathrm{d}l\mathrm{d}\theta\\
&=p_{\underline\theta}+\int_0^{\overline{L}}\left[1-g_{\underline\theta}\left(H\left(l\right)\right)\right]\frac{\partial R_{\underline\theta}(l)}{\partial l}\mathrm{d}l-\mathbb{E}[L]\\
&-\int_{\underline{\theta}}^{\overline{\theta}}\left(\int_0^{\overline{L}}\frac{\partial R_{\theta}(l)}{\partial l}\left(H(l)-g_\theta\left(H\left(l\right)\right)+\frac{\overline{F}(\theta)}{f(\theta)}\frac{\partial g_\theta\left(H\left(l\right)\right)}{\partial \theta}\right)\mathrm{d}l\right)f(\theta)\mathrm{d}\theta\\
&=p_{\underline\theta}+\int_0^{\overline{L}}\left[1-g_{\underline\theta}\left(H\left(l\right)\right)\right]\frac{\partial R_{\underline\theta}(l)}{\partial l}\mathrm{d}l-\mathbb{E}[L]-\int_{\underline{\theta}}^{\overline{\theta}}\left(\int_0^{\overline{L}}\frac{\partial R_{\theta}(l)}{\partial l}J_\theta(l)\mathrm{d}l\right)f(\theta)\mathrm{d}\theta,
\end{align*}
where $J_\theta(l):=H(l)-g_\theta\left(H\left(l\right)\right)+\frac{\overline{F}(\theta)}{f(\theta)}\frac{\partial g_\theta\left(H\left(l\right)\right)}{\partial \theta}$.
\end{proof*}

\subsection*{Proof of Proposition \ref{submodular IC}:}\vspace{-0.25cm}

\begin{proof*}
The ``only if'' part follows directly from Proposition \ref{IC menu}. For the other direction, note first that for any $\theta<\theta'$, we have 
\begin{align*}
&U_{\theta}\left(R_{\theta'},p_{\theta'}\right)
=-p_{\theta'}-\int_0^{\overline{L}}\left[1-g_\theta\left(H\left(l\right)\right)\right]\frac{\partial R_{\theta'}(l)}{\partial l}\mathrm{d}l\\
&=-p_{\underline\theta}-\int_0^{\overline{L}}\left[1-g_{\underline\theta}\left(H\left(l\right)\right)\right]\frac{\partial R_{\underline\theta}(l)}{\partial l}\mathrm{d}l+\int^{\theta'}_{\underline{\theta}}\int_0^{\overline{L}}\frac{\partial g_s\left(H\left(l\right)\right)}{\partial s}\frac{\partial R_{s}(l)}{\partial l}\mathrm{d}l\mathrm{d}s+\int_0^{\overline{L}}\left[g_\theta-g_{\theta'}\left(H\left(l\right)\right)\right]\frac{\partial R_{\theta'}(l)}{\partial l}\mathrm{d}l\\
&=-p_{\underline\theta}-\int_0^{\overline{L}}\left[1-g_{\underline\theta}\left(H\left(l\right)\right)\right]\frac{\partial R_{\underline\theta}(l)}{\partial l}\mathrm{d}l+\int^{\theta}_{\underline{\theta}}\int_0^{\overline{L}}\frac{\partial g_s\left(H\left(l\right)\right)}{\partial s}\frac{\partial R_{s}(l)}{\partial l}\mathrm{d}l\mathrm{d}s+\int_{\theta}^{\theta'}\int_0^{\overline{L}}\frac{\partial g_s\left(H\left(l\right)\right)}{\partial s}\frac{\partial R_{s}(l)}{\partial l}\mathrm{d}l\mathrm{d}s\\
&-\int_0^{\overline{L}}\int^{\theta'}_{\theta}\frac{\partial g_s\left(H\left(l\right)\right)}{\partial s}\mathrm{d}s\frac{\partial R_{\theta'}(l)}{\partial l}\mathrm{d}l\\
&=U_{\theta}\left(R_{\theta},p_{\theta}\right)+\int^{{\theta}'}_{\theta}\int_0^{\overline{L}}\frac{\partial g_s\left(H\left(l\right)\right)}{\partial s}\frac{\partial R_{s}(l)}{\partial l}\mathrm{d}l\mathrm{d}s-\int^{\theta'}_{\theta}\int_0^{\overline{L}}\frac{\partial g_s\left(H\left(l\right)\right)}{\partial s}\frac{\partial R_{\theta'}(l)}{\partial l}\mathrm{d}s\mathrm{d}l\\
&\leq U_{\theta}\left(R_{\theta},p_{\theta}\right),
\end{align*}
where the second equality follows from the expression for $t_{\theta'}$ in \eqref{IC t}, while the fourth equality is based on the expression for $U_\theta\left(R_{\theta},p_{\theta}\right)$ in \eqref{IC utility}, when $p_\theta$ satisfies \eqref{IC t}. Lastly, the inequality follows from  Assumption \ref{gtheta} and the fact that $\frac{\partial R_{\theta}(l)}{\partial l}\leq \frac{\partial R_{\theta'}(l)}{\partial l}$ for any $\theta<\theta'$ given that $R$ is submodular. Similarly, we can show that $U_{\theta'}\left(R_{\theta},p_{\theta}\right)\leq U_{\theta'}\left(R_{\theta'},p_{\theta'}\right).$ Thus, the menu is IC.
\end{proof*}

\subsection*{Proof of Proposition \ref{IR lowest}:}\vspace{-0.25cm}

\begin{proof*}
Consider a  menu $\left(R_{\theta},p_{\theta}\right)_{\theta\in\Theta}\in\cIc.$ By Proposition \ref{IC menu}, we have that 
   \begin{align*}
\frac{\partial U_{\theta}(L,0)}{\partial \theta}=\int_0^{\overline{L}}\frac{\partial g_\theta\left(H\left(l\right)\right)}{\partial \theta}\mathrm{d}l\leq \int_0^{\overline{L}}\frac{\partial g_\theta\left(H\left(l\right)\right)}{\partial \theta}\frac{\partial R_{\theta}(l)}{\partial l}\mathrm{d}l=\frac{\partial U_{\theta}(R_{\theta}, p_\theta)}{\partial \theta},
\end{align*}
where the inequality follows from Assumption \ref{gtheta}, and the fact that $R_\theta\in\mathcal{R}_L.$
Assume that for the lowest type $\underline{\theta},$ the contract is IR. Then, 
$U_{\underline\theta}(R_{\underline\theta},p_{\underline\theta})\geq U_{\underline{\theta}}(L,0).$ For any other $\theta\in \Theta,$ we have that 
\begin{align*}
U_{\theta}(L,0)&=U_{\underline{\theta}}(L,0)+\int_{\underline{\theta}}^\theta \frac{\partial U_{s}(L,0)}{\partial s}\mathrm{d}s\leq U_{\underline{\theta}}(R_{\underline{\theta}}, p_{\underline{\theta}})+\int_{\underline{\theta}}^\theta \frac{\partial U_{s}(R_{s}, p_s)}{\partial s}\mathrm{d}s=U_{\theta}(R_{\theta}, p_\theta),
\end{align*}
which implies that $\left(R_{\theta}, p_\theta\right)_{\theta\in\Theta}\in \cIr$. 
\end{proof*}

\subsection*{Proof of Corollary \ref{IC t interval}:}\vspace{-0.25cm}

\begin{proof*}
First, suppose that $\{R_\theta\}_{\theta\in\Theta}$ is submodular and that  $\{p_\theta\}_{\theta\in\Theta}$ satisfies \eqref{IC t}. By Proposition \ref{submodular IC}, it follows that the menu $(R_\theta,p_\theta)_{\theta\in\Theta}\in\cIc.$ Additionally, if \[p_{\underline\theta}\leq \int_0^{\overline{L}}\left[1-g_{\underline\theta}(H(l))\right]\left(1-\frac{\partial R_{\underline\theta}(l)}{\partial l}\right)\mathrm{d}l,\] then we have 
\begin{align*}
U_{\underline\theta}(R_{\underline\theta},p_{\underline\theta})
&=-p_{\underline\theta}-\int_0^{\overline{L}}\left[1-g_{\underline\theta}(H(l))\right]\frac{\partial R_{\underline\theta}(l)}{\partial l}\mathrm{d}l\\
&\geq-\int_0^{\overline{L}}\left[1-g_{\underline\theta}(H(l))\right]\left(1-\frac{\partial R_{\underline\theta}(l)}{\partial l}\right)\mathrm{d}l-\int_0^{\overline{L}}\left[1-g_{\underline\theta}(H(l))\right]\frac{\partial R_{\underline\theta}(l)}{\partial l}\mathrm{d}l\\
&=-\int_0^{\overline{L}}\left[1-g_{\underline\theta}(H(l))\right]\mathrm{d}l=U_{\underline\theta}(L,0).
\end{align*}

\noindent Hence,  $(R_{\underline\theta},p_{\underline\theta})\in\cIr.$ By Proposition \ref{IR lowest}, it follows that $(R_\theta,p_\theta)_{\theta\in\Theta}\in\cIr.$

\medskip

Conversely, suppose that $\{R_\theta\}_{\theta\in\Theta}$ is submodular and that  $(R_\theta,p_\theta)_{\theta\in\Theta}\in\cIr\cap\cIc.$ By Proposition \ref{submodular IC}, $\{p_\theta\}_{\theta\in\Theta}$ satisfies \eqref{IC t}. Furthermore, since $(R_{\underline\theta}, p_{\underline\theta}) \in \cIr$,  we obtain $U_{\underline\theta}(R_{\underline\theta},p_{\underline\theta})\geq U_{\underline\theta}(L,0),$ which implies that
$$p_{\underline\theta}\leq \int_0^{\overline{L}}\left[1-g_{\underline\theta}(H(l))\right]\left(1-\frac{\partial R_{\underline\theta}(l)}{\partial l}\right)\mathrm{d}l.$$
\end{proof*}

\subsection*{Proof of Proposition \ref{one}:}\vspace{0cm}

\noindent The following result shows that the constraint in Problem \eqref{max2b} is binding at an optimum.

\begin{lemma}
\label{BCmax2b}
If $(R^*,p^*)  \in \mathcal{R}_L \times \bbR_+$ is optimal for Problem \eqref{max2b}, then
\begin{equation}
\label{EqBindingPrem}
p^* = \int_{0}^{ {\overline{L}}}\left[1-g_{\theta_0}\left(H(l)\right)\right] \, \left[1-(R^*)^\prime(l)\right]\, \mathrm{d}l.
\end{equation}
\end{lemma}

\begin{proof}
Suppose that $(R^*,p^*)  \in \mathcal{R}_L \times \bbR_+$ is optimal for Problem \eqref{max2b}, but that
$$0 \leq p^* < \int_{0}^{+\infty}\left[1-g_{\theta_0}\left(H(l)\right)\right] \, \left[1-(R^*)^\prime(l)\right]\, \mathrm{d}l := \tilde p,$$

\noindent Then, for the contract $(R^*, \tilde p)$, we have 
\begin{align*}
U_{\theta_0}(R^*, \tilde p)
&=-\tilde p -\int_{0}^{+\infty}\left[1-g_{\theta_0}\left(H(l)\right)\right] \, (R^*)^\prime(l) \, \mathrm{d}l \\
&=-\int_{0}^{+\infty}\left[1-g_{\theta_0}\left(H(l)\right)\right] \, \left[1-(R^*)^\prime(l)\right]\, \mathrm{d}l -\int_{0}^{+\infty}\left[1-g_{\theta_0}\left(H(l)\right)\right] \, (R^*)^\prime(l) \, \mathrm{d}l \\
&=-\int_{0}^{+\infty}\left[1-g_{\theta_0}\left(H(l)\right)\right] \, \mathrm{d}l \\
&=U_{\theta_0}(L,0),
\end{align*}

\noindent so that the contract $(R^*, \tilde p)$ is feasible for Problem \eqref{max2b}. Moreover,
\begin{align*}
\pi(R^*, \tilde p)
&=\tilde p-\mathbb{E}[L]+\int_0^{\overline{L}}\left[1-H\left(l\right)\right] \, (R^*)^\prime(l)\,\mathrm{d}l\\
&> p^*-\mathbb{E}[L]+\int_0^{\overline{L}}\left[1-H\left(l\right)\right] \, (R^*)^\prime(l)\,\mathrm{d}l\\
&=\pi(R^*,p^*),
\end{align*}

\noindent contradicting the optimality of $(R^*,p^*)$ for Problem \eqref{max2b}. Hence, 
$$p^* = \int_{0}^{+\infty}\left[1-g_{\theta_0}\left(H(l)\right)\right] \, \left[1-(R^*)^\prime(l)\right]\, \mathrm{d}l.$$
\end{proof}

\bigskip
\noindent\underline{Direct Proof of Proposition \ref{one}:}
\begin{proof*}
By Lemma \ref{BCmax2b},  a contract $(R^*,p^*)$ is optimal for Problem \eqref{max2b} if and only if $R^*$ is optimal for 
\begin{equation}
\label{max2c}
\underset{R \in \mathcal{R}_L} \sup \, 
\Bigg\{
\aleph_{\theta_0}(R) := \int_{0}^{\overline L}\left[1-g_{\theta_0}\left(H(l)\right)\right] \, \left[1-R^\prime(l)\right]\, \mathrm{d}l
-\mathbb{E}[L]
+\int_0^{\overline{L}}\left[1-H\left(l\right)\right] \, R^\prime(l)\,\mathrm{d}l
\Bigg\},
\end{equation}

\noindent and 
$$p^* := \int_{0}^{+\infty}\left[1-g_{\theta_0}\left(H(l)\right)\right] \, \left[1-R^\prime(l)\right]\, \mathrm{d}l
=\int_{0}^{\overline L}\left[1-g_{\theta_0}\left(H(l)\right)\right] \, \left[1-R^\prime(l)\right]\, \mathrm{d}l.$$ 

\noindent Now, for each $R \in \mathcal{R}_L$,
\begin{align*}
\aleph_{\theta_0}(R)
&=-\mathbb{E}[L]
+\int_{0}^{\overline L}\left[1-g_{\theta_0}\left(H(l)\right)\right] \, \left[1-R^\prime(l)\right]\, \mathrm{d}l
+\int_0^{\overline{L}}\left[1-H\left(l\right)\right] \, R^\prime(l)\,\mathrm{d}l\\
&=-\mathbb{E}[L]
+\int_{0}^{\overline L}\left[1-g_{\theta_0}\left(H(l)\right)\right] \, \mathrm{d}l
-\int_0^{\overline{L}}\left[H(l) - g_{\theta_0}\left(H(l)\right)\right] \, R^\prime(l)\,\mathrm{d}l\\
&=-\int_{0}^{\overline L}\left[1-H(l)\right] \, \mathrm{d}l
+\int_{0}^{\overline L}\left[1-g_{\theta_0}\left(H(l)\right)\right] \, \mathrm{d}l
-\int_0^{\overline{L}}\left[H(l) - g_{\theta_0}\left(H(l)\right)\right] \, R^\prime(l)\,\mathrm{d}l\\
&=\int_{0}^{\overline L}\left[H(l)-g_{\theta_0}\left(H(l)\right)\right] \, \mathrm{d}l
-\int_0^{\overline{L}}\left[H(l) - g_{\theta_0}\left(H(l)\right)\right] \, R^\prime(l)\,\mathrm{d}l\\
&=\int_{0}^{\overline L}\left[H(l)-g_{\theta_0}\left(H(l)\right)\right] \, \left[1-R^\prime(l)\right] \,\mathrm{d}l.
\end{align*}

\noindent Since $H(l) - g_{\theta_0}\left(H(l)\right) \geq 0$, for all $l \in [0,\overline L]$ (by Assumption \ref{distortion}), and since $R^\prime \in [0,1]$ for all $R \in \mathcal{R}_L$, it follows that the $(R^*)^\prime \equiv 0$ is optimal for Problem \ref{max2c}. Hence, the contract $\left(0,\int_{0}^{\overline L}\left[1-g_{\theta_0}\left(H(l)\right)\right] \, \mathrm{d}l\right) = \left(0,-U_{\theta_0}(L,0)\right)$ is optimal for Problem \eqref{max2b}. Moreover, by \eqref{Ugamma},
$$U_{\theta_0}\left(0,-U_{\theta_0}(L,0)\right) 
=U_{\theta_0}(L,0),$$
meaning that the consumer is indifferent between purchasing insurance and not doing so.
\end{proof*}

\bigskip
{\noindent\underline{Indirect Proof of Proposition \ref{one}:}}
\begin{proof*}
Suppose that $\theta$ is uniformly distributed on the interval $[\theta_0,\theta_0+\epsilon],$ for some constant $\theta_0,$ and a small $\epsilon>0.$ The reciprocal of the hazard rate function of $\theta$ is given by
\[\frac{\overline{F}(\theta)}{f(\theta)}=\frac{\frac{\theta_0+\epsilon-\theta}{\epsilon}}{\frac{1}{\epsilon}}=\theta_0+\epsilon-\theta, \quad\forall \theta\in[\theta_0,\theta_0+\epsilon].\] 
Now, suppose that the distortion function satisfies $g_\theta\equiv g_{\theta_0}$ for all ${\theta\in[\theta_0,\theta_0+\epsilon]}$. Then, the expression in \eqref{virtual value J} simplifies to
  \begin{align*}
      J_\theta(l)&=H(l)-g_\theta\left(H\left(l\right)\right)+\frac{\overline{F}(\theta)}{f(\theta)}\frac{\partial g_\theta\left(H\left(l\right)\right)}{\partial \theta}\\
      &=H(l)-g_\theta\left(H\left(l\right)\right)+(\theta_0+\epsilon-\theta)\cdot 0=H(l)-g_\theta\left(H\left(l\right)\right).
  \end{align*}
Since this function is increasing in $\theta$ for all $l\in[0,\overline{L}],$  and is always non-negative, Theorem \ref{profit solution} implies that the optimal solution satisfies
    \begin{equation*}
\frac{\partial R^\ast_\theta(l)}{\partial l}=0,\quad \text{for}\,\, l\in[0,\overline{L}]\,\text{and}\,\, \theta\in[\theta_0,\theta_0+\epsilon].
\end{equation*}
As $\epsilon$ approaches zero, the risk type parameter $\theta$ converges to a single value, corresponding to a market with only one type of consumer. In this case, the optimal retention function simplifies to  $R_{\theta_0}^\ast(l)=0$ for $l\in[0,\overline{L}]$. For the premium, applying \eqref{optimal t} yields 
$$p_{\theta_0}^\ast=\int_{0}^{\overline{L}}\left[1-g_{\theta_0}\left(H\left(l\right)\right)\right]\mathrm{d}l.$$

Additionally,
$$U_{\theta_0}(R^*_{\theta_0}, p^*_{\theta_0})
=-p^*_{\theta_0} - \int_{0}^{\overline{L}}\left[1-g_{\theta_0}\left(H\left(l\right)\right)\right] \, \frac{\partial R^\ast_{\theta_0}(l)}{\partial l} \, \mathrm{d}l
= -p^*_{\theta_0}
=U_{\theta_0}(L,0).$$
\end{proof*}

\subsection*{Proof of Theorem \ref{profit solution}:}\vspace{-0.25cm}

\begin{proof*}
First note that the profit in \eqref{IC profit} is an increasing function of $p_{\underline{\theta}}.$ Therefore, at the optimum, $p_{\underline{\theta}}$ must take its largest value, provided the IR condition is still satisfied. By Proposition \ref{IRmenu}, we conclude that $p^\ast_{\underline{\theta}}=\int_0^{\overline{L}}\left[1-g_{\underline\theta}(H(l))\right]\left(1-\frac{\partial R_{\underline\theta}(l)}{\partial l}\right)\mathrm{d}l$, where $R_{\underline{\theta}}$ is the retention function for the lowest type consumer.

For $R_\theta\in\mathcal{R}_L$, we know that  $R_{\theta}(0)=0,$ and $\frac{\partial R_{\theta}(l)}{\partial l}\in[0,1]$ for any $\theta\in\Theta.$ An optimal solution for  Problem \ref{max} exists because of the continuity of the profit functional and the compactness of the set of retention functions (Remark \ref{CompactI}). 

With $p_{\underline{\theta}}:=\int_0^{\overline{L}}\left[1-g_{\underline\theta}(H(l))\right]\left(1-\frac{\partial R_{\underline\theta}(l)}{\partial l}\right)\mathrm{d}l$, the insurer's expected profit in \eqref{IC profit} becomes \begin{align}\label{sufficient IC profit} \int_0^{\overline{L}}\left[1-g_{\underline\theta}\left(H\left(l\right)\right)\right]\mathrm{d}l
-\mathbb{E}[L]-\int_{\underline{\theta}}^{\overline{\theta}}\left(\int_0^{\overline{L}}J_\theta(l)\frac{\partial R_\theta(l)}{\partial l}\mathrm{d}l\right)f(\theta)\mathrm{d}\theta. \end{align}

\noindent We maximize \eqref{sufficient IC profit} pointwise. Specifically, for a fixed $\theta\in\Theta,$ we look for 
\begin{align*}
R^\ast_\theta\in\underset{R_\theta\in\mathcal{R}_L}\argmax \ -\int_0^{\overline{L}}J_\theta(l)\frac{\partial R_{\theta}(l)}{\partial l}\mathrm{d}l.
\end{align*}
The maximum is achieved when the retention function takes the form given in \eqref{band-band}.
For any $\theta<\theta',$ since 
 $J_\theta(l)$ is increasing in $\theta$ for all $l,$ it follows that  $J_\theta(l)\leq J_{\theta'}(l)$. Thus, the pointwise maximization solution satisfies
\[\frac{\partial R^\ast_{\theta'}(l)}{\partial l}\leq \frac{\partial R^\ast_\theta(l)}{\partial l}\]
for all $l,$ meaning  $R^\ast_\theta(l)$ is  submodular. Therefore, by Proposition \ref{submodular IC}, it is IC. Additionally, the optimal contract for the lowest risk type  $(R^\ast_{\underline\theta},p^\ast_{\underline\theta})$ satisfies Corollary  \ref{IC t interval},
which confirms that the menu is IR.
\end{proof*}

\subsection*{Proof of Proposition \ref{ComparisionPooling}:}\vspace{-0.25cm}

\begin{proof*}
Let $(R_p, p_p)$ denote the single contract offered to all consumers, where 
$R_p$
  represents the retention and $t_p$	
  the premium. For an consumer of type $\theta$, the utility under this contract is given by
\begin{align*}
U_\theta\left(R_p, p_p\right)
    &= -p_p - \int_0^{\overline{L}}\left[1 - g_\theta\left(H\left(l\right)\right)\right]\frac{\partial R_p(l)}{\partial l} \, \mathrm{d}l.
\end{align*}
With a single contract, one only needs to verify that the IR condition holds.  Specifically, for each $\theta\in\Theta,$ the contract $(R_p, p_p)$ must satisfy $ U_\theta\left(R_p, p_p\right) \geq U_\theta\left(L, 0\right),$ or equivalently,
\begin{align*}
\int_0^{\overline{L}}\left[1 - g_\theta\left(H\left(l\right)\right)\right]\left(1 - \frac{\partial R_p(l)}{\partial l}\right)\mathrm{d}l \geq p_p,
\end{align*}
for any $\theta \in \Theta$. This condition holds for all $\theta \in \Theta$ if and only if it holds for $\underline{\theta}$, as $g_\theta$ is decreasing in $\theta$. Furthermore, since the profit is increasing in $p_p$, the optimal  premium $p_p$ for any given $R_p$ satisfies:
\begin{align*}
    \int_0^{\overline{L}}\left[1 - g_{\underline{\theta}}\left(H\left(l\right)\right)\right]\left(1 - \frac{\partial R_p(l)}{\partial l}\right)\mathrm{d}l = p_p.
\end{align*}
Thus, the insurer's expected profit with the above premium is given by
\begin{align*}
V\left(R_p, p_p\right)
&= \int_\theta \left(p_p - \mathbb{E}[L] + \int_0^{\overline{L}}\left(1 - H\left(l\right)\right)\frac{\partial R_p(l)}{\partial l} \, \mathrm{d}l\right)\mathrm{d}F(\theta) \\
&= \int_0^{\overline{L}}\left[1 - g_{\underline{\theta}}\left(H\left(l\right)\right)\right]\left(1 - \frac{\partial R_p(l)}{\partial l}\right)\mathrm{d}l - \int_0^{\overline{L}}\left[1 - H\left(l\right)\right]\left(1 - \frac{\partial R_p(l)}{\partial l}\right)\mathrm{d}l \\
&= \int_0^{\overline{L}}\left[H\left(l\right) - g_{\underline{\theta}}\left(H\left(l\right)\right)\right]\left(1 - \frac{\partial R_p(l)}{\partial l}\right)\mathrm{d}l.
\end{align*}

\noindent Hence, the profit decreases as the marginal retention function increases. Thus, at the optimum, the retention function satisfies $R_p^\ast = 0 $. Consequently, the maximum profit with a single contract is,
\begin{align*}
V\left(R_p^\ast, p_p^\ast\right) 
= \int_0^{\overline{L}}\left[H\left(l\right) - g_{\underline{\theta}}\left(H\left(l\right)\right)\right]\mathrm{d}l,
\end{align*}

\noindent and
\begin{align*}
V\left(R_p^\ast, p_p^\ast\right)
&\leq 
\int_0^{\overline{L}}\left[H(l) - g_{\underline\theta}\left(H\left(l\right)\right)\right]\mathrm{d}l - \int_{\underline{\theta}}^{\overline{\theta}}\left(\int_0^{\overline{L}} J_\theta(l)\One_{\{J_\theta(l) < 0\}}\mathrm{d}l\right)f(\theta) \, \mathrm{d}\theta \\
&=-U_{\underline{\theta}}\left(L, 0\right) - \mathbb{E}[L] - \int_{\underline{\theta}}^{\overline{\theta}}\left(\int_0^{\overline{L}} J_\theta(l)\One_{\{J_\theta(l) < 0\}}\mathrm{d}l\right)f(\theta) \, \mathrm{d}\theta,
\end{align*}
which is the expected profit obtained from the solution given in Theorem \ref{profit solution}. Thus, for consumers with varying risk attitudes, designing a menu of contracts to achieve a separating equilibrium is always more advantageous for the insurer than providing a uniform contract that leads to a pooling equilibrium.
\end{proof*}

\subsection*{Proof of Proposition \ref{menu properties}:}\vspace{-0.35cm}

\begin{proof*}
\begin{enumerate}
\item Following from Theorem \ref{profit solution}, $R^\ast_\theta$ is decreasing in $\theta$ due to submodularity. Substituting the optimal retention into the premium expression in \eqref{IC t} and taking the derivative yields
\begin{align*}
\frac{\partial p^\ast_\theta}{\partial \theta}=-\int_0^{\overline{L}}\frac{\partial g_\theta\left(H\left(l\right)\right)}{\partial s}\frac{\partial R^\ast_{\theta}(l)}{\partial l}\mathrm{d}l+\int_0^{\overline{L}}\frac{\partial g_\theta\left(H\left(l\right)\right)}{\partial s}\frac{\partial R^\ast_{\theta}(l)}{\partial l}\mathrm{d}l-\int_0^{\overline{L}}\left(1-g_\theta\left(H\left(l\right)\right)\right)\frac{\partial}{\partial \theta}\frac{\partial R^\ast_{\theta}(l)}{\partial l}\mathrm{d}l\geq0,
\end{align*} 
since $\frac{\partial}{\partial \theta}\frac{\partial R^\ast_{\theta}(l)}{\partial l}\leq0.$

\item This follows from the fact that $J(l,\overline{\theta})\leq 0$ for all $l.$ 

\item This follows from the expression for $p^\ast_{\underline\theta}$ in \eqref{optimal t}.

\item From the expression of the consumer's utility for an IC menu given in \eqref{IC utility}, 
\begin{align*}
\frac{\partial U_\theta\left(R^\ast_{\theta},p^\ast_{\theta}\right)}{\partial \theta}
&=\int_0^{\overline{L}}\frac{\partial g_\theta\left(H\left(l\right)\right)}{\partial \theta}\frac{\partial R^\ast_{\theta}(l)}{\partial l}\mathrm{d}l\leq 0,
\end{align*}
since $g_\theta$ satisfies Assumption \ref{gtheta}. In addition, the second derivative is given by
\begin{align*}
\frac{\partial^2 U_\theta\left(R^\ast_{\theta},p^\ast_{\theta}\right)}{\partial \theta^2}
&=\int_0^{\overline{L}}\frac{\partial}{\partial \theta}\left(\frac{\partial g_\theta\left(H\left(l\right)\right)}{\partial \theta}\right)\cdot\frac{\partial R^\ast_{\theta}(l)}{\partial l}\mathrm{d}l+\int_0^{\overline{L}}\frac{\partial g_\theta\left(H\left(l\right)\right)}{\partial \theta}\cdot \frac{\partial }{\partial \theta}\left(\frac{\partial R^\ast_{\theta}(l)}{\partial l}\right)\mathrm{d}l\geq 0,
\end{align*}
if $\frac{\partial g_\theta(s)}{\partial \theta}$ is increasing. Therefore, the utility of wealth at an optimum is decreasing and convex in $\theta$.

\item The profit in \eqref{pi} for the contract $(R_\theta^\ast,p_\theta^\ast)$ is
\begin{align}\label{profit gamma}
&\pi\left(R^\ast_{\theta},p^\ast_{\theta}\right)=p^\ast_\theta-\mathbb{E}[L]+\int_0^{\overline{L}}\left[1-H\left(l\right)\right]\frac{\partial R^\ast_{\theta}(l)}{\partial l}\mathrm{d}l\nonumber\\
&=\int_0^{\overline{L}}\left[1-g_{\underline\theta}(H(l))\right]\mathrm{d}l-\int^\theta_{\underline{\theta}}\int_0^{\overline{L}}\frac{\partial g_s\left(H\left(l\right)\right)}{\partial s}\frac{\partial R^\ast_{s}(l)}{\partial l}\mathrm{d}l\mathrm{d}s-\int_0^{\overline{L}}\left[1-g_\theta\left(H\left(l\right)\right)\right]\frac{\partial R^\ast_{\theta}(l)}{\partial l}\mathrm{d}l-\mathbb{E}[L]\nonumber\\
&+\int_0^{\overline{L}}\left[1-H\left(l\right)\right]\frac{\partial R^\ast_{\theta}(l)}{\partial l}\mathrm{d}l\nonumber\\
&=\int_0^{\overline{L}}\left[H\left(l\right)-g_{\underline\theta}\left(H\left(l\right)\right)\right]\mathrm{d}l-\int_0^{\overline{L}}\left[H\left(l\right)-g_\theta\left(H\left(l\right)\right)\right]\frac{\partial R^\ast_{\theta}(l)}{\partial l}\mathrm{d}l
-\int_{\underline{\theta}}^\theta\int_0^{\overline{L}}\frac{\partial g_s\left(H\left(l\right)\right)}{\partial s}\frac{\partial R^\ast_{s}(l)}{\partial l}\mathrm{d}l\mathrm{d}s.
\end{align}
Taking the derivative with respect to $\theta$ yields  
\begin{align*}
\frac{\partial \pi\left(R^\ast_{\theta},p^\ast_{\theta}\right)}{\partial \theta}&=
-\int_0^{\overline{L}}\frac{\partial g_\theta\left(H\left(l\right)\right)}{\partial \theta}\frac{\partial R^\ast_{\theta}(l)}{\partial l}\mathrm{d}l+\int_0^{\overline{L}}\frac{\partial g_\theta\left(H\left(l\right)\right)}{\partial \theta}\frac{\partial R^\ast_{\theta}(l)}{\partial l}\mathrm{d}l\\
&-\int_0^{\overline{L}}\left[H\left(l\right)- g_\theta\left(H\left(l\right)\right)\right]\frac{\partial }{\partial \theta}\frac{\partial R^\ast_{\theta}(l)}{\partial l}\mathrm{d}l\\
&=-\int_0^{\overline{L}}\left[H\left(l\right)- g_\theta\left(H\left(l\right)\right)\right]\frac{\partial }{\partial \theta}\frac{\partial R^\ast_{\theta}(l)}{\partial l}\mathrm{d}l\geq0,
\end{align*}
 for any $\theta\in\Theta$.
\item At an optimum, the expected profit in \eqref{IC profit} is 
\begin{align*}
V\left(\left(R^\ast_{\theta}, p^\ast_{\theta}\right)_{\theta\in\Theta}\right)
&=\int_0^{\overline{L}}\left[H(l)-g_{\underline\theta}\left(H\left(l\right)\right)\right]\mathrm{d}l-\int_{\underline{\theta}}^{\overline{\theta}}\left(\int_0^{\overline{L}}\frac{\partial R^\ast_{\theta}(l)}{\partial l}J_\theta(l)\mathrm{d}l\right)f(\theta)\mathrm{d}\theta\\
&=\int_0^{\overline{L}}\left[H(l)-g_{\underline\theta}\left(H\left(l\right)\right)\right]\mathrm{d}l-\int_{\underline{\theta}}^{\overline{\theta}}\left(\int_0^{\overline{L}}J_\theta(l)\One_{\{J_\theta(l)<0\}}\mathrm{d}l\right)f(\theta)\mathrm{d}\theta\\
&\geq\int_0^{\overline{L}}\left[H(l)-g_{\underline\theta}\left(H\left(l\right)\right)\right]\mathrm{d}l=-\mathbb{E}[L]-U_{\underline\theta}(L,0)>0,
\end{align*}
where the last inequality follows from Proposition \ref{riskaverse g}(1). This inequality is strict since $\underline{\theta}$ is risk averse. Thus, the insurer earns a positive profit.
\end{enumerate}
\end{proof*}

\vspace{-0.5cm}

\subsection*{Proof of Theorem \ref{special case}:}\vspace{-0.25cm}

\begin{proof*}
By Theorem \ref{profit solution}, for any $\theta<\overline{\theta},$ we have that an optimal solution satisfies $\frac{\partial R_\theta^\ast(l)}{\partial l}=0$ for $J_\theta(l)\geq0,$ and $\frac{\partial R_\theta^\ast(l)}{\partial l}=1$ for $J_\theta(l)<0.$ When there exists a unique $l_\theta\in [0,\overline{L}]$ such that $J_\theta(l)\leq 0$ for $l\leq l_\theta$ and $J_\theta(l)\geq0$ for $l\geq l_\theta,$ we obtain
\begin{equation*}
R_\theta^\ast(l)=
\left\{
\begin{array}[c]{ll}%
l, & \textit{if} \quad l\leq l_\theta,\vspace{0.2cm}\\
l_\theta, & \textit{if} \quad l> l_\theta.
\end{array}
\right.  
\end{equation*}

In contract, if for each $\theta<\overline{\theta}$, there exists a unique $l_\theta\in[0,\overline{L}]$ such that $J_\theta(l)\geq 0$ for $l\leq l_\theta$ and $J_\theta(l)\leq0$ for $l\geq l_\theta,$ we obtain
    \begin{equation*}
R_\theta^\ast(l)=
\left\{
\begin{array}[c]{ll}%
0, & \textit{if} \quad l\leq l_\theta,\vspace{0.2cm}\\
l-l_\theta, & \textit{if} \quad l> l_\theta.
\end{array}
\right.  
\end{equation*}
\end{proof*}

\vspace{-0.25cm}
\subsection*{Proof of Proposition \ref{profit expression cost}:}\vspace{-0.25cm}

\begin{proof*}
For any IR contract, if $R_{\theta}(l)=l$ for all $l\in[0,\overline{L}]$, we know that $p_\theta=0,$ and thus $\pi(R_\theta,p_\theta)=0.$ Let $\Theta_R$ denote the set of $\theta$ values for which $R_{\theta}(l)=l$ for all $l\in[0,\overline{L}]$, i.e., $\Theta_R=\left\{\theta\in\Theta\mid R_{\theta}(l)=l\,\text{for}\,\, l\in[0,\overline{L}]\right\}.$ By the definition of $\tilde\pi$ in \eqref{tilde pi}, we have
\begin{align*}
\tilde V\left(\left(R_{\theta}, p_{\theta}\right)_{\theta\in\Theta}\right)
&=\int_{\Theta /\Theta_R}\left(\pi(R_\theta,p_\theta)-k\right)\mathrm{d}F(\theta)\\
&=\int_{\Theta }\left(\pi(R_\theta,p_\theta)-k\right)\mathrm{d}F(\theta)-\int_{\Theta_R}\left(\pi(R_\theta,p_\theta)-k\right)\mathrm{d}F(\theta)\\
&=V\left(\left(R_{\theta}, p_{\theta}\right)_{\theta\in\Theta}\right)-k-\int_{\Theta_R}\left(0-k\right)\mathrm{d}F(\theta)\\
&=V\left(\left(R_{\theta}, p_{\theta}\right)_{\theta\in\Theta}\right)-k+k\cdot\int_{\theta}\One_{\left\{\theta\in\Theta_R\right\}}\mathrm{d}F(\theta)\\
&=V\left(\left(R_{\theta}, p_{\theta}\right)_{\theta\in\Theta}\right)-k\cdot\mathbb{P}\left(\theta\notin\Theta_R\right).
\end{align*}
\end{proof*}

\vspace{-0.5cm}
\subsection*{Proof of Theorem \ref{profit solution cost}:}\vspace{-0.25cm}

\begin{proof*}
By Proposition \ref{profit expression cost}(1), the menu that maximizes the insurer profit when cost is included in \eqref{VRT cost} also solves the following problem
\begin{align}\label{cost problem}
\max\limits_{R\in\mathcal{R}_L} \quad \int_0^{\overline{L}}\left[1-g_{\underline\theta}(H(l))\right]\mathrm{d}l
-\mathbb{E}[L]-\int_{\underline{\theta}}^{\overline{\theta}}\left(\int_0^{\overline{L}}J_\theta(l)\frac{\partial R_\theta(l)}{\partial l}\mathrm{d}l\right)f(\theta)\mathrm{d}\theta-k+k\cdot\mathbb{P}\left(\theta\in\Theta_R\right).
\end{align}
We solve problem \eqref{cost problem} pointwise. Specifically, for a fixed $\theta\in\Theta,$ we look for $R_\theta^\ast$ such that
\begin{align*}
R_\theta^\ast\in\argmax\limits_{R\in\mathcal{R}_L} \quad-\left(\int_0^{\overline{L}}J_\theta(l)\frac{\partial R_\theta(l)}{\partial l}\mathrm{d}l\right) +k\cdot\One_{\left\{\theta\in\Theta_R\right\}},
\end{align*}
or
\[R_\theta^\ast\in\argmax\limits_{R\in\mathcal{R}_L} \quad-\left(\int_0^{\overline{L}}J_\theta(l)\frac{\partial R_\theta(l)}{\partial l}\mathrm{d}l\right) +k\cdot\One_{\left\{R_\theta(l)\equiv l\right\}}.\]
If $\One_{\left\{R_\theta(l)\equiv l\right\}}=0,$  at the optimum, we know that $\frac{\partial R^\ast_\theta(l)}{\partial l}=1$ when $J_\theta(l)<0$, and  $\frac{\partial R^\ast_\theta(l)}{\partial l}=0$ when $J_\theta(l)>0.$ Hence the total profit is $-\int_0^{\overline{L}}J_\theta(l)\One_{\left\{J_\theta(l)<0\right\}}\mathrm{d}l.$  If $\One_{\left\{R_\theta(l)\equiv l\right\}}=1$,
 the profit is $-\int_0^{\overline{L}}J_\theta(l)\mathrm{d}l+k.$

If $-\int_0^{\overline{L}}J_\theta(l)\One_{\left\{J_\theta(l)<0\right\}}\mathrm{d}l\geq-\int_0^{\overline{L}}J_\theta(l)\mathrm{d}l+k,$ or equivalently,  $k\leq\int_0^{\overline{L}}J_\theta(l)\One_{\left\{J_\theta(l)>0\right\}}\mathrm{d}l,$ then it remains profitable to provide some insurance even with a fixed cost. However, if $k>\int_0^{\overline{L}}J_\theta(l)\One_{\left\{J_\theta(l)>0\right\}}\mathrm{d}l,$  the fixed cost exceeds the benefits collected, resulting in no insurance being provided. In this case, the optimal retention function is given by $R^\ast_\theta(l)=l$, for all $l$.

Additionally, the integral $\int_0^{\overline{L}}J_\theta(l)\One_{\left\{J_\theta(l)>0\right\}}\mathrm{d}l$ is an increasing function of $\theta.$ This is because, for $\theta_1<\theta_2,$ we know that $J_{\theta_1}(l)\leq J_{\theta_2}(l)$ for any $l\in [0,\overline{L}]$, by assumption. Thus,
$\One_{\left\{J_{\theta_1}(l)>0\right\}}\leq \One_{\left\{J_{\theta_2}(l)>0\right\}}.$ Then
\[\int_0^{\overline{L}}J_{\theta_1}(l)\One_{\left\{J_{\theta_1}(l)>0\right\}}\mathrm{d}l\leq \int_0^{\overline{L}}J_{\theta_2}(l)\One_{\left\{J_{\theta_1}(l)>0\right\}}\mathrm{d}l\leq \int_0^{\overline{L}}J_{\theta_2}(l)\One_{\left\{J_{\theta_2}(l)>0\right\}}\mathrm{d}l.\]

\noindent Thus:
\begin{enumerate}
\item If $\int_0^{\overline{L}}J_{\overline\theta}(l)\One_{\left\{J_{\overline\theta}(l)>0\right\}}\mathrm{d}l=\int_0^{\overline{L}}H(l)-g_{\overline\theta}\left(H\left(l\right)\right)\mathrm{d}l<k,$  it is unprofitable for the insurer to trade with the highest-risk type, and thus it becomes unprofitable for all risk types. Consequently, the insurer chooses not to offer any contracts.

\item Otherwise,  let $\theta^\ast=\max\left\{\theta\in\Theta\mid\int_0^{\overline{L}}J_\theta(l)\One_{\left\{J_\theta(l)>0\right\}}\mathrm{d}l\leq k\right\}.$ For any $\theta\leq \theta^\ast,$ we have $R_\theta^\ast(l)=l$ for $l\in [0,\overline{L}].$ When $\theta>\theta^\ast,$ $R_\theta^\ast$ takes the form given in \eqref{band-band}, which is submodular as shown by Theorem \ref{profit solution}. Therefore, the solution to  \eqref{cost problem} is also submodular and thus satisfies the IC condition. The IR condition follows similarly to the proof of Theorem \ref{profit solution}, and is omitted here.
\end{enumerate}

Next, we compare the separating equilibrium contract with the pooling equilibrium. Let $(R_p, p_p)$ denote the single contract offered to all consumers. By Proposition \ref{ComparisionPooling},  the maximum profit with a single contract is
\begin{align*}
\tilde V\left(R_p^\ast, p_p^\ast\right) 
= \int_0^{\overline{L}}\left[H\left(l\right) - g_{\underline{\theta}}\left(H\left(l\right)\right)\right]\mathrm{d}l-k.
\end{align*}

\noindent For the separating menu of contracts, the insurer's profit at an optimum is 
\begin{align*}
\tilde V\left(\left(R^\ast_{\theta}, p^\ast_{\theta}\right)_{\theta\in\Theta}\right)
&= \int_0^{\overline{L}}\left[H(l) - g_{\underline\theta}\left(H\left(l\right)\right)\right]\mathrm{d}l - \int_{\underline{\theta}}^{\overline{\theta}}\left(\int_0^{\overline{L}} J_{\theta}(l)\One_{\{J_{\theta}(l) < 0\}}\mathrm{d}l\right)f(\theta) \, \mathrm{d}\theta -k\overline{F}\left(\theta^\ast\right).
\end{align*}

\noindent Therefore,
\begin{align}
\tilde V\left(\left(R^\ast_{\theta}, p^\ast_{\theta}\right)_{\theta\in\Theta}\right)
-\tilde V\left(R_p^\ast, p_p^\ast\right)
=- \int_{\underline{\theta}}^{\overline{\theta}}\left(\int_0^{\overline{L}} J_{\theta}(l)\One_{\{J_{\theta}(l) < 0\}}\mathrm{d}l\right)f(\theta) \, \mathrm{d}\theta+kF\left(\theta^\ast\right)\geq0,
\end{align}
which demonstrates the optimality of separating insurance.
\end{proof*}

\subsection*{Proof of Proposition \ref{ProbWFmaxPO}:}\vspace{-0.25cm}

\begin{proof*}
Suppose that there exists a probability measure $\eta$ on $\left(\Theta, \Sigma_\mathcal{B}(\Theta)\right)$ such that   $\mu$ is absolutely continuous with respect to $\eta$,  and some $\alpha \in (0,1)$, such that $(R^\ast_\theta,p^\ast_\theta)_{\theta\in\Theta}$ is optimal for \eqref{POmax2}, but that $(R^\ast_\theta,p^\ast_\theta)_{\theta\in\Theta} \notin \mathcal{IPO}$. Then there exists $\left(R_\theta,p_\theta\right)_{\theta\in\Theta}\in\cIr\cap\cIc$ such that
\begin{align*} 
U_\theta\left(R_{\theta},p_{\theta}\right)
\geq
U_\theta\left(R^\ast_{\theta},p^\ast_{\theta}\right), \ \forall \, \theta \in \Theta,
\quad \text{and} \quad 
\int_\Theta
\pi\left(R_{\theta},p_{\theta}\right)\mathrm{d}\mu \geq \int_\Theta
\pi\left(R^\ast_{\theta},p^\ast_{\theta}\right)\mathrm{d}\mu, \end{align*}
with at least one strict inequality. Moreover, if $\displaystyle\int_\Theta
\pi\left(R_{\theta},p_{\theta}\right)\mathrm{d}\mu = \int_\Theta
\pi\left(R^\ast_{\theta},p^\ast_{\theta}\right)\mathrm{d}\mu$, then 
$$\mu \bigg(\Big\{\theta \in \Theta: U_{\theta}\left(R_{\theta},p_{\theta}\right)
>
U_{\theta}\left(R^\ast_{\theta},p^\ast_{\theta}\right)\Big\}\bigg) > 0,$$
and hence
$$\eta \bigg(\Big\{\theta \in \Theta: U_{\theta}\left(R_{\theta},p_{\theta}\right)
>
U_{\theta}\left(R^\ast_{\theta},p^\ast_{\theta}\right)\Big\}\bigg) > 0.$$
Therefore, 
$$\alpha \, \int_\Theta U_\theta\left(R_{\theta},p_{\theta}\right) \, \mathrm{d}\eta
+
(1-\alpha) \, 
\int_\Theta
\pi\left(R_{\theta},p_{\theta}\right)\,\mathrm{d}\mu
>
\alpha \, \int_\Theta U_\theta\left(R^\ast_{\theta},p^\ast_{\theta}\right) \, \mathrm{d}\eta
+
(1-\alpha) \, 
\int_\Theta
\pi\left(R^\ast_{\theta},p^\ast_{\theta}\right)\,\mathrm{d}\mu,$$
contradicting the optimality of $(R^\ast_\theta,p^\ast_\theta)_{\theta\in\Theta}$ for \eqref{POmax2}. Hence, $(R^\ast_\theta,p^\ast_\theta)_{\theta\in\Theta} \in \mathcal{IPO}$.
\end{proof*}

\subsection*{Proof of Theorem \ref{profit solution pareto}:}
\vspace{-0.25cm}

\begin{proof*} 
Form \eqref{U expression} and \eqref{pi}, it follows that the social welfare function can be rewritten as
\begin{align*}
W_{\eta,\alpha}\left(\left(R_{\theta}, p_{\theta}\right)_{\theta\in\Theta}\right) 
&= 
\alpha \,
\int_{\underline\theta}^{\overline\theta} U_\theta\left(R_{\theta},p_{\theta}\right)  \, \mathrm{d}F_\eta(\theta)
+
(1-\alpha) \,
\int_{\underline\theta}^{\overline\theta}
\pi\left(R_{\theta},p_{\theta}\right)\,\mathrm{d}F(\theta)\\
&=\alpha\int_{\underline\theta}^{\overline\theta} \bigg(-p_\theta-\int_0^{\overline{L}}\left[1-g_\theta(H(l))\right]\frac{\partial R_\theta(l)}{\partial l}\mathrm{d}l\bigg)  \, \mathrm{d}F_\eta(\theta)\\
&\qquad\qquad+(1-\alpha)\,
\int_{\underline\theta}^{\overline\theta}
\left(p_\theta-\mathbb{E}[L]+\int_0^{\overline{L}}\left[1-H(l)\right]\frac{\partial R_\theta(l)}{\partial l}\mathrm{d}l\right)\,\mathrm{d}F(\theta)\\
&=\int_{\underline\theta}^{\overline\theta}\int_0^{\overline{L}}\left((1-\alpha)[1-H(l)]-\alpha\frac{\mathrm{d}F_\eta(\theta)}{\mathrm{d}F(\theta)}[1-g_\theta(H(l))]\right)\frac{\partial R_\theta(l)}{\partial l}\mathrm{d}l\,\mathrm{d}F(\theta)\\
&\qquad\qquad+\int_{\underline\theta}^{\overline\theta}\left(1-\alpha-\alpha\frac{\mathrm{d}F_\eta(\theta)}{\mathrm{d}F(\theta)}\right)p_\theta\,\mathrm{d}F(\theta)-(1-\alpha)\mathbb{E}[L].
\end{align*}

\noindent Now, for every $\left(R_\theta,p_\theta\right)_{\theta\in\Theta}\in\cIc$, it follows from Proposition \ref{IC menu} that the premium satisfies \eqref{IC t}. Substituting this premium structure into the social welfare function yields
\begin{align}\label{Walpha}
   & W_{\eta,\alpha}\left(\left(R_{\theta}, p_{\theta}\right)_{\theta\in\Theta}\right)
    = \int_{\underline\theta}^{\overline\theta}\int_0^{\overline{L}}\left((1-\alpha)[1-H(l)]-\alpha\frac{\mathrm{d}F_\eta(\theta)}{\mathrm{d}F(\theta)}[1-g_\theta(H(l))]\right)\frac{\partial R_\theta(l)}{\partial l}\mathrm{d}l\,\mathrm{d}F(\theta)\nonumber\\
    &+\int_{\underline\theta}^{\overline\theta}\left(1-\alpha-\alpha\frac{\mathrm{d}F_\eta(\theta)}{\mathrm{d}F(\theta)}\right)\left(p_{\underline\theta}+\int_0^{\overline{L}}\left[1-g_{\underline\theta}\left(H\left(l\right)\right)\right]\frac{\partial R_{\underline\theta}(l)}{\partial l}\mathrm{d}l-\,\int_0^{\overline{L}}\left[1-g_\theta\left(H\left(l\right)\right)\right]\frac{\partial R_{\theta}(l)}{\partial l}\mathrm{d}l\right)\,\mathrm{d}F(\theta)\nonumber\\
    &-\int_{\underline\theta}^{\overline\theta}\left(1-\alpha-\alpha\frac{\mathrm{d}F_\eta(\theta)}{\mathrm{d}F(\theta)}\right)\left(\int^\theta_{\underline{\theta}}\int_0^{\overline{L}} \, \frac{\partial g_s\left(H\left(l\right)\right)}{\partial s}\frac{\partial R_{s}(l)}{\partial l}\mathrm{d}l\mathrm{d}s\right)\,\mathrm{d}{F}(\theta)-(1-\alpha)\mathbb{E}[L]\nonumber\\  &=\left(p_{\underline\theta}+\int_0^{\overline{L}}\left[1-g_{\underline\theta}\left(H\left(l\right)\right)\right]\frac{\partial R_{\underline\theta}(l)}{\partial l}\mathrm{d}l\right)\int_{\underline\theta}^{\overline\theta}\left(1-\alpha-\alpha\frac{\mathrm{d}F_\eta(\theta)}{\mathrm{d}F(\theta)}\right)\,\mathrm{d}F(\theta)-(1-\alpha)\mathbb{E}[L]\nonumber\\
    &-\left(1-\alpha\right)\left(\int_{\underline\theta}^{\overline\theta}\int_0^{\overline{L}}\left[H(l)-g_{\theta}\left(H\left(l\right)\right)\right]\frac{\partial R_{\theta}(l)}{\partial l}\mathrm{d}l\,\mathrm{d}F(\theta)+\int_{\underline\theta}^{\overline\theta}\int^\theta_{\underline{\theta}}\int_0^{\overline{L}} \, \frac{\partial g_s\left(H\left(l\right)\right)}{\partial s}\frac{\partial R_{s}(l)}{\partial l}\mathrm{d}l\mathrm{d}s\,\mathrm{d}{F}(\theta)\right)\nonumber\\
&+\alpha\int_{\underline\theta}^{\overline\theta}\int^\theta_{\underline{\theta}}\int_0^{\overline{L}} \, \frac{\partial g_s\left(H\left(l\right)\right)}{\partial s}\frac{\partial R_{s}(l)}{\partial l}\mathrm{d}l\mathrm{d}s\,\mathrm{d}{F}_\eta(\theta)\nonumber\\ 
&=(1-2\alpha)\left(p_{\underline\theta}+\int_0^{\overline{L}}\left[1-g_{\underline\theta}\left(H\left(l\right)\right)\right]\frac{\partial R_{\underline\theta}(l)}{\partial l}\mathrm{d}l\right)-(1-\alpha)\mathbb{E}[L]\nonumber\\
&-(1-\alpha)\int_{\underline\theta}^{\overline\theta}\int_0^{\overline{L}}\left[H(l)-g_{\theta}\left(H\left(l\right)\right)\right]\frac{\partial R_{\theta}(l)}{\partial l}\mathrm{d}l\,\mathrm{d}F(\theta)\nonumber\\
    &+\left.\left(1-\alpha\right)\left(\int^\theta_{\underline{\theta}}\int_0^{\overline{L}} \, \frac{\partial g_s\left(H\left(l\right)\right)}{\partial s}\frac{\partial R_{s}(l)}{\partial l}\mathrm{d}l\mathrm{d}s\right)\overline{F}(\theta)\right\vert_{\theta=\underline{\theta}}^{\theta=\overline{\theta}}-\left(1-\alpha\right)\int_{\underline\theta}^{\overline\theta}\int_0^{\overline{L}} \, \frac{\partial g_\theta\left(H\left(l\right)\right)}{\partial\theta}\frac{\partial R_{\theta}(l)}{\partial l}\mathrm{d}l\,\overline{F}(\theta)\mathrm{d}\theta\nonumber\\
    &-\left.\alpha\left(\int^\theta_{\underline{\theta}}\int_0^{\overline{L}} \, \frac{\partial g_s\left(H\left(l\right)\right)}{\partial s}\frac{\partial R_{s}(l)}{\partial l}\mathrm{d}l\mathrm{d}s\right)\overline{F}_\eta(\theta)\right\vert_{\theta=\underline{\theta}}^{\theta=\overline{\theta}}+\alpha\int_{\underline\theta}^{\overline\theta}\int_0^{\overline{L}} \, \frac{\partial g_\theta\left(H\left(l\right)\right)}{\partial\theta}\frac{\partial R_{\theta}(l)}{\partial l}\mathrm{d}l\,\overline{F}_\eta(\theta)\mathrm{d}\theta\nonumber\\
&=(1-2\alpha)\left(p_{\underline\theta}+\int_0^{\overline{L}}\left[1-g_{\underline\theta}\left(H\left(l\right)\right)\right]\frac{\partial R_{\underline\theta}(l)}{\partial l}\mathrm{d}l\right)-(1-\alpha)\mathbb{E}[L]\nonumber\\
   &-\int_{\underline\theta}^{\overline\theta}\int_0^{\overline{L}}\left(\left(1-\alpha\right)\left[H(l)-g_{\theta}\left(H\left(l\right)\right)\right]+\frac{(1-\alpha)\overline{F}(\theta)-\alpha\overline{F}_\eta(\theta)}{f(\theta)}\frac{\partial g_\theta\left(H\left(l\right)\right)}{\partial\theta}\right)\frac{\partial R_{\theta}(l)}{\partial l}\mathrm{d}lf(\theta)\,\mathrm{d}\theta\nonumber\\  
   &=(1-2\alpha)\left(p_{\underline\theta}+\int_0^{\overline{L}}\left[1-g_{\underline\theta}\left(H\left(l\right)\right)\right]\frac{\partial R_{\underline\theta}(l)}{\partial l}\mathrm{d}l\right)-(1-\alpha)\mathbb{E}[L]-\int_{\underline\theta}^{\overline\theta}\int_0^{\overline{L}}J_{\eta,\alpha,\theta}(l)\frac{\partial R_{\theta}(l)}{\partial l}\mathrm{d}lf(\theta)\,\mathrm{d}\theta,
\end{align}
where
\begin{align}\label{JEtaAlphaTh}
J_{\eta,\alpha,\theta}(l)
&:=\left(1-\alpha\right)\left[H(l)-g_{\theta}\left(H\left(l\right)\right)\right]+\frac{(1-\alpha)\overline{F}(\theta)-\alpha\overline{F}_\eta(\theta)}{f(\theta)}\frac{\partial g_\theta\left(H\left(l\right)\right)}{\partial\theta}\nonumber\\
&=\left(1-\alpha\right)\left[H(l)-g_{\theta}\left(H\left(l\right)\right)\right]+\frac{\overline{F}(\theta)}{f(\theta)}\left(1-\alpha-\alpha\frac{\overline{F}_\eta(\theta)}{\overline{F}(\theta)}\right)\frac{\partial g_\theta\left(H\left(l\right)\right)}{\partial\theta}.
\end{align}

\begin{enumerate}
\item If $\alpha\leq\frac{1}{2},$ the social welfare function \eqref{Walpha} is a non-decreasing  function of $p_{\underline{\theta}}.$ Therefore, at the optimum, $p_{\underline{\theta}}$ must take its largest value, provided the IR condition is still satisfied. By Proposition \ref{IRmenu}, we conclude that $p^\ast_{\underline{\theta}}=\int_0^{\overline{L}}\left[1-g_{\underline\theta}(H(l))\right]\left(1-\frac{\partial R_{\underline\theta}(l)}{\partial l}\right)\mathrm{d}l$, where $R_{\underline{\theta}}$ is the retention function for the lowest type consumer.
With $p_{\underline{\theta}}:=\int_0^{\overline{L}}\left[1-g_{\underline\theta}(H(l))\right]\left(1-\frac{\partial R_{\underline\theta}(l)}{\partial l}\right)\mathrm{d}l$, the social welfare function simplifies to:
\begin{align*} (1-2\alpha)\int_0^{\overline{L}}\left[1-g_{\underline\theta}\left(H\left(l\right)\right)\right]\mathrm{d}l
-(1-\alpha)\mathbb{E}[L]-\int_{\underline{\theta}}^{\overline{\theta}}\left(\int_0^{\overline{L}}J_{\eta,\alpha,\theta}(l)\frac{\partial R_\theta(l)}{\partial l}\mathrm{d}l\right)f(\theta)\mathrm{d}\theta. \end{align*}
We maximize this function pointwise, following the approach in Theorem \ref{profit solution}. First, we analyze the function $J_{\eta,\alpha,\theta}(l)$ in \eqref{JEtaAlphaTh}. The first term, $\left(1-\alpha\right)\left[H(l)-g_{\theta}\left(H\left(l\right)\right)\right]\geq0$ by Assumption \ref{distortion}. For the second term, it is evident that $\frac{\overline{F}(\theta)}{f(\theta)}\frac{\partial g_\theta\left(H\left(l\right)\right)}{\partial\theta}\leq0,$ since $g_\theta$ satisfies Assumption \ref{gtheta}. For the remaining part, we observe that $\left(1-\alpha-\alpha\frac{\overline{F}_\eta(\theta)}{\overline{F}(\theta)}\right)\bigg\vert_{\theta=\underline\theta}=1-2\alpha.$ Furthermore, at $\theta=\overline\theta,$ using Remark \ref{EtaAndMu}, we obtain $$\left(1-\alpha-\alpha\frac{\overline{F}_\eta(\theta)}{\overline{F}(\theta)}\right)\bigg\vert_{\theta=\overline\theta}
=1-\alpha-\alpha\frac{f_\eta(\overline{\theta})}{f(\overline{\theta})}.$$

\noindent Since $\frac{\overline{F}_\eta(\theta)}{\overline{F}(\theta)} $ is non-decreasing in $\theta$, it follows that the function 
\[\theta \mapsto 1-\alpha-\alpha\frac{\overline{F}_\eta(\theta)}{\overline{F}(\theta)}\]
is non-increasing. Therefore, if $1-\alpha-\alpha\frac{f_\eta(\overline{\theta})}{f(\overline{\theta})}>0,$ or equivalently, $\alpha<\frac{f(\overline{\theta})}{f_\eta(\overline{\theta})+f(\overline{\theta})},$ then
$$\frac{\overline{F}(\theta)}{f(\theta)}\left(1-\alpha-\alpha\frac{\overline{F}_\eta(\theta)}{\overline{F}(\theta)}\right)\frac{\partial g_\theta\left(H\left(l\right)\right)}{\partial\theta}\leq0.$$ 

\noindent For a fixed $\theta\in\Theta,$ the optimal retention solves the following maximization problem:
\begin{align*}
R^\ast_{\eta,\alpha,\theta}\in\underset{R_{\eta,\alpha,\theta}\in\mathcal{R}_L}\argmax \ -\int_0^{\overline{L}}J_{\eta,\alpha,\theta}(l)\frac{\partial R_{\eta,\alpha,\theta}(l)}{\partial l}\mathrm{d}l.
\end{align*}

\noindent Hence, the optimal retention function satisfies \eqref{band-band pareto}. When $\alpha$ and $\eta$ are specified, 
for any $\theta<\theta',$ and if
 $J_{\eta,\alpha,\theta}(l)$ is non-decreasing in $\theta$ for all $l,$ it follows that  $J_{\eta,\alpha,\theta}(l)\leq J_{\eta,\alpha,\theta'}(l)$. Thus, the pointwise maximization solution satisfies
\[\frac{\partial R^\ast_{\eta,\alpha,\theta'}(l)}{\partial l}\leq \frac{\partial R^\ast_{\eta,\alpha,\theta}(l)}{\partial l}\]
for all $l,$ implying that $R^\ast_{\eta,\alpha,\theta}(l)$ is  submodular. Therefore, by Proposition \ref{submodular IC}, it satisfies the IC condition. Additionally, the optimal contract for the lowest risk type  $(R^\ast_{\eta,\alpha,\underline\theta},p^\ast_{\eta,\alpha,\underline\theta})$ satisfies Corollary  \ref{IC t interval},
ensuring that the menu is IR. 

\medskip

\item If $1-2\alpha\geq0$ and $1-\alpha-\alpha\frac{f_\eta(\overline{\theta})}{f(\overline{\theta})}\leq0,$ or equivalently, $\alpha\in\left[\frac{f(\overline{\theta})}{f_\eta(\overline{\theta})+f(\overline{\theta})},\frac{1}{2}\right],$ then there exists $\theta_\alpha\in\Theta$ such that $1-\alpha-\alpha\frac{\overline{F}_\eta(\theta_0)}{\overline{F}(\theta_0)}=0.$  Additionally, $1-\alpha-\alpha\frac{\overline{F}_\eta(\theta)}{\overline{F}(\theta)}\geq0$   when $\theta<\theta_\alpha$, and $1-\alpha-\alpha\frac{\overline{F}_\eta(\theta)}{\overline{F}(\theta)}\leq0$   when $\theta>\theta_\alpha.$ 
Therefore, if the function $J_{\eta,\alpha,\theta}(l)$ is non-decreasing in $\theta$ when $\theta<\theta_\alpha,$ the optimal retention function $R^\ast_{\eta,\alpha,\theta}$ follows the form given in \eqref{band-band pareto}. When $\theta\geq\theta_\alpha,$ since $J_{\eta,\alpha,\theta}(l)\geq0$ for all $l,$ it follows that $R^\ast_{\eta,\alpha,\theta}=0.$

\medskip

\item If $\alpha>\frac{1}{2},$ the social welfare function \eqref{Walpha} is a non-increasing function of $p_{\eta,\alpha,\underline\theta}.$ Therefore, at the optimum, $p_{\eta,\alpha,\underline\theta}$ must take its smallest value while still satisfying the IR condition. By Proposition \ref{IRmenu}, we obtain $p^\ast_{\eta,\alpha,\underline\theta}=0.$  Substituting this into the social welfare function \eqref{Walpha}, we obtain
\begin{align*} &(1-2\alpha)\int_0^{\overline{L}}\left[1-g_{\underline\theta}\left(H\left(l\right)\right)\right]\frac{\partial R_{\eta,\alpha,\underline\theta}(l)}{\partial l}\mathrm{d}l
-(1-\alpha)\mathbb{E}[L]-\int_{\underline{\theta}}^{\overline{\theta}}\left(\int_0^{\overline{L}}J_{\theta,\alpha,\eta}(l)\frac{\partial R_{\eta,\alpha,\theta}(l)}{\partial l}\mathrm{d}l\right)f(\theta)\mathrm{d}\theta\\
&=\int_{\underline{\theta}}^{\overline{\theta}}\int_0^{\overline{L}}\left((1-2\alpha)\left[1-g_{\underline\theta}\left(H\left(l\right)\right)\right]\frac{\partial R_{\eta,\alpha,\underline\theta}(l)}{\partial l}-J_{\eta,\alpha,\theta}(l)\frac{\partial R_{\eta,\alpha,\theta}(l)}{\partial l}\right)\mathrm{d}lf(\theta)\mathrm{d}\theta-(1-\alpha)\mathbb{E}[L].
\end{align*}
In this case, $1-\alpha-\alpha\frac{\overline{F}_\eta(\theta)}{\overline{F}(\theta)}<0$, for all $\theta$. Consequently,  $J_{\eta,\alpha,\theta}(l)\geq0$, for all $\theta.$
 For $\theta=\underline\theta,$ we define
\begin{align*}
  \tilde J_{\eta,\alpha,\underline\theta}(l):=-(1-2\alpha)\left[1-g_{\underline\theta}\left(H\left(l\right)\right)\right]+J_{\eta,\alpha,\underline\theta}(l)\geq J_{\eta,\alpha,\underline\theta}(l)\geq0.
\end{align*}
 The optimal retention function for the lowest risk type satisfies
\begin{align*} 
R_{\eta,\alpha,\underline\theta}^\ast\in\underset{R_{\eta,\alpha,\underline\theta}\in\mathcal{R}_L}\argmax \ -\int_0^{\overline{L}}\tilde J_{\eta,\alpha,\underline\theta}(l)\frac{\partial R_{\eta,\alpha,\underline \theta}(l)}{\partial l}\mathrm{d}l.
\end{align*}
For all other risk types $\theta$, the optimal retention function satisfies
\begin{align*} 
R_{\eta,\alpha,\theta}^\ast\in\underset{R_{\eta,\alpha,\theta}\in\mathcal{R}_L}\argmax \ -\int_0^{\overline{L}}J_{\eta,\alpha,\theta}(l)\frac{\partial R_{\eta,\alpha, \theta}(l)}{\partial l}\mathrm{d}l,
\end{align*}
The pointwise maximization implies that $\frac{\partial R^\ast_{\eta,\alpha,\theta}(l)}{\partial l}=0$ for all $l,$ and thus $R_{\eta,\alpha,\theta}^\ast\equiv 0$ for any $\theta\in\Theta.$
\end{enumerate}
\vspace{-0.4cm}
\end{proof*}

\subsection*{Proof of Proposition \ref{profit solution pareto properties}:}\vspace{-0.45cm}

\begin{proof*}
\begin{enumerate}    
\item By \eqref{J and J alpha}, if $J_{\eta,\alpha,\theta}(l)<0$ then  $J_\theta(l)<0$, since $\frac{\partial g_\theta\left(H\left(l\right)\right)}{\partial \theta}\leq0.$
Now, suppose there exist two weight parameters, $\alpha_1$ and $\alpha_2$ such that $\alpha_1< \alpha_2.$ Suppose, by way of contradiction, that    $$\frac{\partial R_{\eta,\alpha_1,\theta}^{\ast}(l)}{\partial l}< \frac{\partial R_{\eta,\alpha_2,\theta}^{\ast}(l)}{\partial l}$$ for some  $l\in[0,\overline{L}]$. Since the right-hand side of the inequality is at its maximum, it follows that  $\frac{\partial R_{\eta,\alpha_2,\theta}^{\ast}(l)}{\partial l}=1.$ This implies that $J_{\eta,\alpha_2,\theta}(l)<0$, which in turn leads to $J_\theta(l)<0.$ Given that $\alpha_1< \alpha_2,$ we obtain
$$(1-\alpha_1)J_{\theta}(l)<(1-\alpha_2)J_\theta(l),$$ which further implies that
$$J_{\eta,\alpha_1,\theta}(l)<J_{\eta,\alpha_2,\theta}(l)<0.$$ 
Consequently, we must also have $\frac{\partial R_{\eta,\alpha_1,\theta}^{\ast}(l)}{\partial l}=1,$ contradicting our initial assumption. Therefore, we conclude that $\frac{\partial R_{\eta,\alpha_1,\theta}^{\ast}(l)}{\partial l}\geq \frac{\partial R_{\eta,\alpha_2,\theta}^{\ast}(l)}{\partial l}$ for all $l.$

\medskip

\item The utility function for a type-$\theta$ consumer under the contract $(R_{\eta,\alpha,\theta}^{\ast},p^\ast_{\eta,\alpha,\theta})$ is given by
\begin{align}
\label{Ualpha}
&U_\theta(R_{\eta,\alpha,\theta}^{\ast},p^\ast_{\eta,\alpha,\theta})
=-p^\ast_{\eta,\alpha,\theta}-\int_0^{\overline{L}}\left[1-g_\theta\left(H\left(l\right)\right)\right]\frac{\partial R^\ast_{\eta,\alpha,\theta}(l)}{\partial l}\mathrm{d}l\nonumber\\
&=-p^\ast_{\eta,\alpha,\underline\theta}-\int_0^{\overline{L}}\left[1-g_{\underline\theta}\left(H\left(l\right)\right)\right]\frac{\partial R^\ast_{\eta,\alpha,\underline\theta}(l)}{\partial l}\mathrm{d}l+\int^\theta_{\underline{\theta}}\int_0^{\overline{L}}\frac{\partial g_s\left(H\left(l\right)\right)}{\partial s}\frac{\partial R^\ast_{\eta,\alpha,s}(l)}{\partial l}\mathrm{d}l\mathrm{d}s\nonumber\\
&=-\int_0^{\overline{L}}\left[1-g_{\underline\theta}\left(H\left(l\right)\right)\right]\mathrm{d}l+\int^\theta_{\underline{\theta}}\int_0^{\overline{L}}\frac{\partial g_s\left(H\left(l\right)\right)}{\partial s}\frac{\partial R^\ast_{\eta,\alpha,s}(l)}{\partial l}\mathrm{d}l\mathrm{d}s.
\end{align}
If $\alpha_1<\alpha_2,$ then
$$U_\theta(R_{\eta,\alpha_1,\theta}^{\ast},p^\ast_{\eta,\alpha_1,\theta})\leq U_\theta(R_{\eta,\alpha_2,\theta}^{\ast},p^\ast_{\eta,\alpha_2,\theta})$$ 
since $\frac{\partial g_s\left(H\left(l\right)\right)}{\partial s}\leq0$, and, by Proposition \ref{profit solution pareto properties}(1), $\frac{\partial R_{\eta,\alpha_1,\theta}^{\ast}(l)}{\partial l}\geq \frac{\partial R_{\eta,\alpha_2,\theta}^{\ast}(l)}{\partial l}$. When $\alpha\geq\frac{1}{2}$, the utility function satisfies $U_\theta(R_{\eta,\alpha,\theta}^{\ast},p^\ast_{\eta,\alpha,\theta})=0$, for every $\theta\in\Theta.$

\medskip

\item For the insurer, when $\alpha\leq\frac{1}{2},$ the premium schedule satisfies \eqref{IC t}, and the total profit is given by
\begin{align}\label{Valpha}
&V\big((R^\ast_{\eta,\alpha,\theta}, p^\ast_{\eta,\alpha,\theta})_{\theta\in\Theta}\big)\nonumber\\
&=p^\ast_{\eta,\alpha,\underline\theta}+\int_0^{\overline{L}}\left[1-g_{\underline\theta}\left(H\left(l\right)\right)\right]\frac{\partial R^\ast_{\eta,\alpha,\underline\theta}(l)}{\partial l}\mathrm{d}l-\mathbb{E}[L]-\int_{\underline{\theta}}^{\overline{\theta}}\left(\int_0^{\overline{L}}J_{\theta}(l)\frac{\partial R^\ast_{\eta,\alpha,\theta}(l)}{\partial l}\mathrm{d}l\right)f(\theta)\mathrm{d}\theta\nonumber\\
&=\int_0^{\overline{L}}\left[1-g_{\underline\theta}\left(H\left(l\right)\right)\right]\mathrm{d}l-\mathbb{E}[L]-\int_{\underline{\theta}}^{\overline{\theta}}\left(\int_0^{\overline{L}}J_{\theta}(l)\frac{\partial R^\ast_{\eta,\alpha,\theta}(l)}{\partial l}\mathrm{d}l\right)f(\theta)\mathrm{d}\theta.%
\end{align} 
 When $\alpha_1<\alpha_2$, we have $\frac{\partial R_{\eta,\alpha_1,\theta}^{\ast}(l)}{\partial l}\geq \frac{\partial R_{\eta,\alpha_2,\theta}^{\ast}(l)}{\partial l}$ by Proposition \ref{profit solution pareto properties}(1). Since, at the optimum, $\frac{\partial R^\ast_{\eta,\alpha,\theta}(l)}{\partial l}=\{0,1\}$, suppose that for some $l,$
 \[1=\frac{\partial R_{\eta,\alpha_1,\theta}^{\ast}(l)}{\partial l}> \frac{\partial R_{\eta,\alpha_2,\theta}^{\ast}(l)}{\partial l}=0.\]
Then, we have $J_{\eta,\alpha_1,\theta}(l)<0,$ which, by \eqref{J and J alpha}, implies that
$J_{\theta}(l)<0$. Thus, it follows that $$-\int_0^{\overline{L}}\frac{\partial R^\ast_{\eta,\alpha_1,\theta}(l)}{\partial l}J_{\theta}(l)\mathrm{d}l\geq -\int_0^{\overline{L}}\frac{\partial R^\ast_{\eta,\alpha_2,\theta}(l)}{\partial l}J_{\theta}(l)\mathrm{d}l.$$
Integrating over $\theta$ in $\Theta,$ we obtain $$V\big((R^\ast_{\eta,\alpha_1,\theta}, p^\ast_{\eta,\alpha_1,\theta})_{\theta\in\Theta}\big)\geq V\big((R^\ast_{\eta,\alpha_2,\theta}, p^\ast_{\eta,\alpha_2,\theta})_{\theta\in\Theta}\big).$$
Additionally, from \eqref{Valpha}, we derive the bound
\begin{align*}
&V\big((R^\ast_{\eta,\alpha,\theta}, p^\ast_{\eta,\alpha,\theta})_{\theta\in\Theta}\big)\geq\int_0^{\overline{L}}\left[1-g_{\underline\theta}\left(H\left(l\right)\right)\right]\mathrm{d}l-\mathbb{E}[L]=\int_0^{\overline{L}}\left[H(l)-g_{\underline\theta}\left(H\left(l\right)\right)\right]\mathrm{d}l>0. 
\end{align*} 
When $\alpha>\frac{1}{2}$, the insurer's total profit is given by
\begin{align*}
V\big((R^\ast_{\eta,\alpha,\theta}, p^\ast_{\eta,\alpha,\theta})_{\theta\in\Theta}\big)
&=\int_{\underline\theta}^{\overline\theta}
\pi\left(R^\ast_{\eta,\alpha,\theta},p^\ast_{\eta,\alpha,\theta}\right)f(\theta)\mathrm{d}\theta\\
&=\int_{\underline\theta}^{\overline\theta}\left(p^\ast_{\eta,\alpha,\theta}-\mathbb{E}\left[L-R^\ast_{\eta,\alpha,\theta}(L)\right]\right)f(\theta)\mathrm{d}\theta
=-\mathbb{E}[L].
\end{align*}
\end{enumerate}
\vspace{-0.6cm}
\end{proof*}

\subsection*{Proof of Corollary \ref{one Pareto}:}\vspace{-0.25cm}

\begin{proof*} If $\Theta=\{\theta_0\},$ the social welfare function is given by:
\begin{align}\label{Walpha one}
&   \alpha\, U_{\theta_0}(R,p)+(1-\alpha)\,\pi(R,p)\nonumber\\
&=\alpha\left(-p_{\theta_0}-\int_0^{\overline{L}}\left[1-g_{\theta_0}\left(H\left(l\right)\right)\right]\frac{\partial R_{\theta_0}(l)}{\partial l}\mathrm{d}l\right)+(1-\alpha)\left(p_{\theta_0}-\mathbb{E}[L]+\int_0^{\overline{L}}\left[1-H\left(l\right)\right]\frac{\partial R_{\theta_0}(l)}{\partial l}\mathrm{d}l\right)\nonumber\\
&=(1-2\alpha)p_{\theta_0}+\int_0^{\overline{L}}\left((1-\alpha)\left[1-H\left(l\right)\right]-\alpha\left[1-g_{\theta_0}\left(H\left(l\right)\right)\right]\right)\frac{\partial R_{\theta_0}(l)}{\partial l}\mathrm{d}l-(1-\alpha)\mathbb{E}[L].
\end{align}

\begin{enumerate}
\item If $\alpha\leq\frac{1}{2}$, we see that the function in \eqref{Walpha one} is a non-decreasing function of 
$p_{\theta_0}$. By Proposition \ref{IRmenu}, we conclude that at the optimum, when 
$R_{\theta_0}\in\mathcal{R}_L$, the premium is
\[p^\ast_{\theta_0} 
=
\int_0^{\overline{L}}\left[1-g_{\theta}(H(l))\right]\left(1-\frac{\partial R_{\theta_0}(l)}{\partial l}\right)\mathrm{d}l.\]
Thus, the social welfare function becomes:
\begin{align*}
    &(1-2\alpha)\int_0^{\overline{L}}\left[1-g_{\theta}(H(l))\right]\left(1-\frac{\partial R_{\theta_0}(l)}{\partial l}\right)\mathrm{d}l\\
    &+\int_0^{\overline{L}}\left((1-\alpha)\left[1-H\left(l\right)\right]-\alpha\left[1-g_{\theta_0}\left(H\left(l\right)\right)\right]\right)\frac{\partial R_{\theta_0}(l)}{\partial l}\mathrm{d}l-(1-\alpha)\mathbb{E}[L]\\
    &=(1-2\alpha)\int_0^{\overline{L}}\left[1-g_{\theta}(H(l))\right]\mathrm{d}l+(1-\alpha)\int_0^{\overline{L}}\left(\left[1-H\left(l\right)\right]-\left[1-g_{\theta_0}\left(H\left(l\right)\right)\right]\right)\frac{\partial R_{\theta_0}(l)}{\partial l}\mathrm{d}l-(1-\alpha)\mathbb{E}[L]\\
    &=(1-2\alpha)\int_0^{\overline{L}}\left[1-g_{\theta}(H(l))\right]\mathrm{d}l+(1-\alpha)\int_0^{\overline{L}}\left[g_{\theta_0}\left(H\left(l\right)\right)-H\left(l\right)\right]\frac{\partial R_{\theta_0}(l)}{\partial l}\mathrm{d}l-(1-\alpha)\mathbb{E}[L].
\end{align*}
The maximizer of this equation is $\frac{\partial R^\ast_{\theta_0}(l)}{\partial l}\equiv 0$, since $g_{\theta_0}\left(H\left(l\right)\right)-H\left(l\right)\leq0$ for all $l.$

\item If $\alpha>\frac{1}{2}$, the function in \eqref{Walpha one} becomes a decreasing function of 
 $p_{\theta_0}.$ Thus, at the optimum, the premium is 
$p^\ast_{\theta_0} 
=
0$.
The social welfare function then becomes:
\begin{align*}
&\int_0^{\overline{L}}\left((1-\alpha)\left[1-H\left(l\right)\right]-\alpha\left[1-g_{\theta_0}\left(H\left(l\right)\right)\right]\right)\frac{\partial R_{\theta_0}(l)}{\partial l}\mathrm{d}l-(1-\alpha)\mathbb{E}[L].
\end{align*}
Since  $1-H\left(l\right)\leq 1-g_{\theta_0}\left(H\left(l\right)\right)$ for all $l,$ and $1-\alpha<\alpha,$ it follows that  $$(1-\alpha)\left[1-H\left(l\right)\right]\leq\alpha\left[1-g_{\theta_0}\left(H\left(l\right)\right)\right].$$ Thus, the maximizer of this equation is again $\frac{\partial R^\ast_{\theta_0}(l)}{\partial l}\equiv 0$.
\end{enumerate}
\vspace{-0.6cm}
\end{proof*}

\vspace{-0.25cm}

\bibliographystyle{ecta}
\setlength{\bibsep}{0.08cm}
\bibliography{ref}

\end{document}